\documentclass{article}

\usepackage{arxiv}

\usepackage{graphicx}
\usepackage{natbib}
\usepackage{doi}
\usepackage{epsfig} 
\usepackage{subcaption}
\usepackage{amsmath,amsfonts} 
\usepackage{bm}

\usepackage{psfrag}

\usepackage{hyperref}

\hypersetup{
  colorlinks   = true, 
  urlcolor     = blue, 
  linkcolor    = blue, 
  citecolor   = red 
}

\usepackage{microtype}
\usepackage{booktabs}
\usepackage[ruled,vlined,linesnumbered]{algorithm2e}

\DeclareMathOperator*{\argmin}{arg\,min}

\usepackage{multirow}
\usepackage{url}
\usepackage{makecell}
\usepackage{relsize}

\usepackage{mathtools} 
\DeclareMathAlphabet{\pazocal}{OMS}{zplm}{m}{n}
\makeatletter
\newcommand{\vast}{\bBigg@{3}}
\makeatother
\makeatletter
\renewcommand*\env@matrix[1][\arraystretch]{%
  \edef\arraystretch{#1}%
  \hskip -\arraycolsep
  \let\@ifnextchar\new@ifnextchar
  \array{*\c@MaxMatrixCols c}}
\makeatother
\usepackage{amssymb}
            
\title{Platoon Fundamental Diagram estimation can be Markovian: evidence from human- and self-driven vehicle trajectories}

\author{Michail A. Makridis\\
	IVT, Civil Environmental and Geomatic Engineering\\
	ETH Zurich, Switzerland\\
	\texttt{mmakridis@ethz.ch} \\
	\And
	Anastasios Kouvelas \\
	IVT, Civil Environmental and Geomatic Engineering\\
	ETH Zurich, Switzerland\\
	\texttt{kouvelas@ethz.ch} \\
 \And
	Jorge A. Laval \\
	Civil and Environmental Engineering\\
	Georgia Institute of Technology, US\\
	\texttt{jorge.laval@ce.gatech.edu} \\
}

\renewcommand{\shorttitle}{Platoon Fundamental Diagram estimation can be Markovian}

\begin{document}
\maketitle

\begin{abstract}
We propose a simple and effective method to derive the Fundamental Diagram (FD) from platoon vehicle trajectories. Average traffic state variables are computed using Edie's generalized definitions within time-dependent trapezoidal space-time areas. To obtain a clear FD, we employ a bivariate data aggregation technique to eliminate scatter. Our findings are as follows:
(i) The proposed method demonstrates a remarkably consistent relation between the traffic variables and a clear triangular shape for autonomously-driven vehicles. 
(ii) The FDs are invariant to several factors of heterogeneity such as the platoon length, vehicle characteristics, road particularities, and data acquisition accuracy.
(iii) ACC-driven vehicle platoons with minimum headway setting achieve much higher capacity, roughly 90\% than those with a large headway setting. 
(iv) Connectivity might increase capacity.
(v) Human drivers have a wider near-capacity operation area, showing different behaviors at high speeds than low ones, and
(vi) Safety concerns might arise due to high values of backward wave speed for ACC-driven vehicles.
Comparative analysis with the state-of-the-art confirms the validity of our approach. The proposed method stands out due to its simplicity and accuracy, which paves the way for practical applications in real-time traffic flow monitoring and control within modern intelligent transportation systems.  
\end{abstract}

\section{Introduction}
\label{sec:introduction}
Traffic investigations, initiated nearly 90 years ago, aimed to better comprehend the intricate phenomena occurring in road transport systems. Among early pioneers was Greenshields who in 1933 noticed a linear relation between speed and traffic density and a parabolic relation between speed and traffic flow using the assumption that traffic flow equals traffic density multiplied by speed, also known as Fundamental Diagram (FD). Since then, numerous studies in the literature exploit the concept of FD as a congestion modeling tool for traffic estimation purposes at link or network levels \cite{kouvelas_enhancing_2017,makridis_adaptive_2023, du_adaptive_2023}.

The study of vehicle platoon trajectories has been ongoing for over 50 years. In 1966, \cite{treiterer_investigation_1975} used aerial photographs from a helicopter, following a specific vehicle and monitoring the traffic around it, as all vehicles progress along the roadway. Subsequently, \cite{treiterer_hysteresis_1974} observed a generally retarded behavior displayed by a platoon of vehicles after emerging from a kinematic disturbance as compared to the behavior of the same vehicles approaching the disturbance, known today as traffic hysteresis phenomenon. This discovery sparked a continued interest in platoon dynamics. 

The exploration of platoon trajectories is appealing due to its ability to offer insights into the behavior of different drivers under similar conditions, which directly influences traffic state. The existence of diverse driver behaviors on public roads has been widely recognized as a catalyst for various traffic phenomena. Consequently, there is a strong interest in accurately modeling and simulating this heterogeneity, as discussed in \cite{makridis_formalizing_2020}. Moreover, the integration of automation and connectivity in vehicles further amplifies the impact of drivers' behaviors on road capacity and utilization \cite{ioannou_automated_1997, van_arem_impact_2006,makridis_impact_2020}. However, there is limited information in the existing literature regarding how different driver modes may contribute to distinct Fundamental Diagrams (FDs).

Fortunately, a growing number of experimental datasets that include human-driven and automated vehicles emerge, as in \cite{makridis_openacc_2021,gunter_are_2021}. Such datasets provide valuable opportunities to gain fresh insights into the topic. While there is extended research on the microscopic properties of partially or fully automated vehicles such as reaction time \cite{brunner_comparing_2022, makridis_response_2020}, string instability \cite{li_car-following_2021, gunter_are_2021}, safety \cite{lu_crash_2021, mattas_fuzzy_2020,mullakkal-babu_probabilistic_2020}, and energy efficiency \cite{apostolakis_energy-based_2023, he_energy_2020, ma_analysis_2021}, there are very few works that explicitly relate their microscopic behavior with the traffic state. Although there are some recent efforts, see \cite{chen_towards_2017,shi_constructing_2021}, achieving generalization is not a straightforward task due to challenges from various factors such as road geometry, platoon length, diverse vehicle specifications, or data acquisition frequency. This paper takes advantage of the operational properties of commercial ACC (and CACC) systems to aim for a constant headway policy. It proposes a simple and robust numerical method to describe macroscopic traffic flow properties of ACC- and CACC-driven platoons of vehicles. 

By analyzing observations from either fully or partially automated vehicles and applying an aggregation technique, we establish a clear relationship between the three traffic flow variables: speed, density, and flow. This results in clear FD for homogeneous drivers in platoons, using only inexpensive raw trajectory observations. Due to the absence of ground-truth information, the validity of the method in terms of traffic state inference is assessed using one of the currently most reliable methods for traffic state inference in the literature \cite{laval_hysteresis_2011,maiti_universality_2023}, and the resemblance is remarkably high. 

Our method proves to be robust against heterogeneity from means of observations, including variations in noise levels, platoon synthesis, platoon length, and traffic conditions. Additionally, it eliminates the need for estimating the backward wave speed and minimizes computational complexity. The results reveal a triangular shape for the FD of partially automated vehicles and provide concrete insights into the expected capacity and speed-dependent dynamic behavior of human drivers compared to partially automated vehicles. At the same time, safety concerns arise for ACC-driven vehicles with minimum distance settings.

In the remainder of this paper, we present in Section~\ref{sec:methodology} the proposed methodology, while Section~\ref{sec:results} discusses the results and finally, conclusions are summarized in Section~\ref{sec:conclusion}.

\section{Methodology}\label{sec:methodology}
First, we adopt the well-known Edie's generalized definitions for raw movement data aggregation to derive fundamental traffic variables, namely, flow, density, and space-mean speed, see \cite{edie_discussion_1965,cassidy_relation_1997}. 

It is important to note that in this procedure, a rectangular area needs to be defined, and the flow, density, and speed are calculated based on the number of vehicle-kilometers traveled and vehicle-time traveled within this region. Based on the most acknowledged measurement method in the literature, see \cite{laval_hysteresis_2011}, the shape of this region should be a sided rectangle including the platoon vehicles with side slopes representing the expected backward wave speed. This approach is crucial to avoid blending different traffic states and to obtain a reliable approximation of the prevailing traffic state within the given space-time area. However, implementation-wise there are numerical difficulties for discrete observations. Additionally, defining a global representation of the fundamental diagram (FD) for a driver mode remains an unresolved matter, particularly when analyzing large trajectory datasets due to the significant variability observed.

In this study, we first attempt to estimate the traffic state at a given time based on two consecutive observations. The data resolution (usually 10$Hz$) does not affect the final results. Through this process, the estimated traffic states are derived, which are equal in number to the total number of observations minus one.

\begin{figure*}[!b]
    \begin{subfigure}[t]{0.45\textwidth}
        \includegraphics[width=\textwidth]{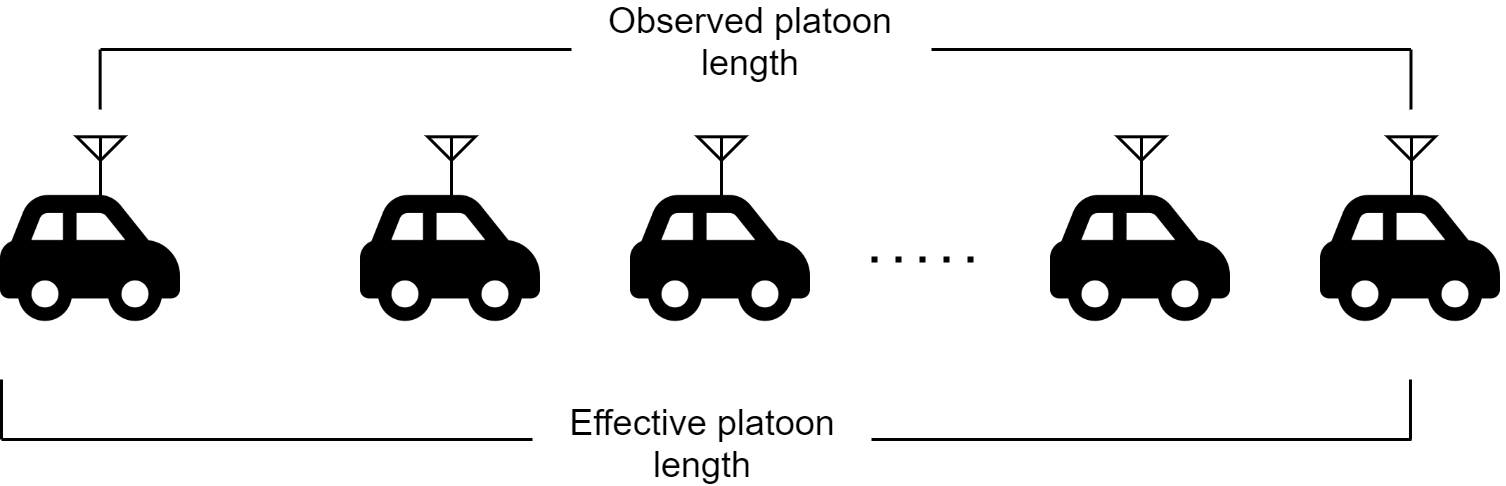}
        \caption{Illustration of the observed and effective platoon length.}\label{fig:platoon}
    \end{subfigure}
    ~ 
    \begin{subfigure}[t]{0.45\textwidth}
        \includegraphics[width=\textwidth]{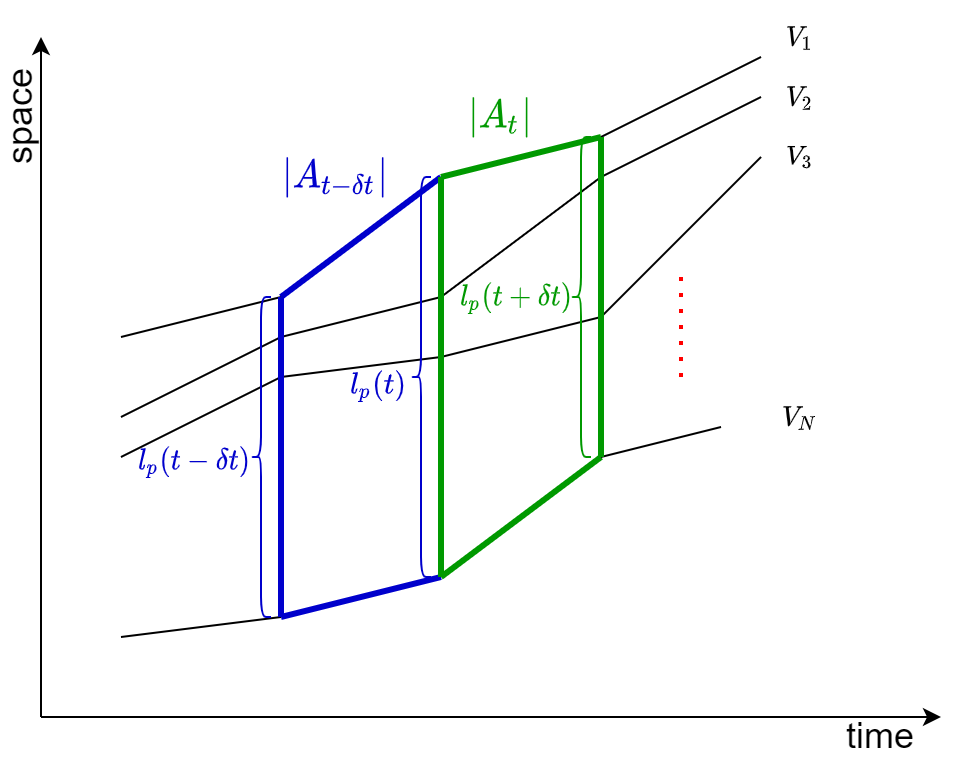}
        \caption{Illustration of moving scalene trapezoid space-time area.}\label{fig:area}
    \end{subfigure}
    \caption{Schematic representation of an observed platoon and a dynamic area for instantaneous traffic state estimation.}
    \label{fig:edies}
\end{figure*}

Here, we present an example that illustrates how to determine the effective platoon length, which establishes the boundaries of our designated area. Figure~\ref{fig:platoon} depicts a platoon of vehicles with their antennas used for data acquisition. The observed platoon length represents the distance between the first and last antenna, excluding a portion of the first and last vehicle. Assuming equal vehicle lengths and identical antenna positions on each vehicle, this corresponds to the distance between the back bumper of the leader and the back bumper of the last follower. To account for assumptions' variability and measurement noise, we incorporate a small buffer space (approximately 3 meters). This adjusted length is referred to as the effective platoon length, which varies dynamically as the vehicles move. Figure~\ref{fig:area} provides a schematic representation of the area used for the computation of speed, density, and flow. For the mathematical representation, we refer the reader to \ref{sec:app_methodology}.

The computation of the instantaneous traffic states described above is carried out for all available datasets, classified according to the driver mode (human, ACC with fixed distance setting, and CACC). This generates a large number of potential states and a significant amount of scatter. We aim to verify if all the observations that correspond to the same driver mode, e.g., ACC with the minimum setting, correspond to the same fundamental diagram. To derive a single fundamental diagram (FD) relationship, we employ a simple and efficient data aggregation technique across all available data.

\begin{figure*}[h]
    \begin{subfigure}[t]{0.65\textwidth}
        \includegraphics[width=\textwidth]{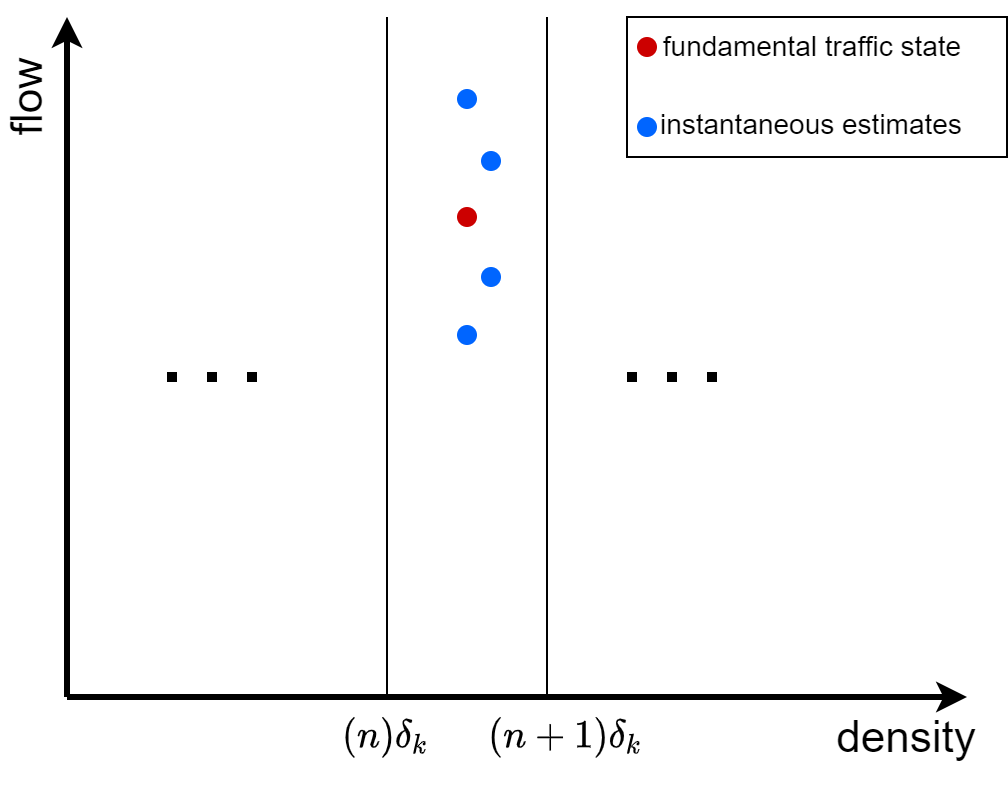}
    \end{subfigure}
    \caption{Schematic representation of the aggregation method. Blue dots correspond to instantaneous traffic state estimates for the same driver mode. Each dot corresponds to an observation without considering the experiment, platoon size, powertrain, etc. All estimates have instantaneous density withing the bin range, i.e., $k \in (n\delta_k, (n+1)\delta_k)$. The inferred fundamental traffic state for the corresponding bin is derived by averaging all observed instantaneous densities. Therefore each density bin corresponds to one flow value.}
    \label{fig:aggregation_process}
\end{figure*}

In the flow-density representation, we divide the observed density range into bins of length $\delta_k$.  Within each density interval, we calculate the average of all observed flow values, yielding a single estimated flow value per density bin. An exemplary illustration is given in Figure~\ref{fig:aggregation_process}. As expected, the accuracy of the fundamental traffic state improves with more observations. The results show that the inferred fundamental traffic states per driver mode are independent of the bin length, $\delta_k$. A similar aggregation process is applied to the observed speed range with aggregation of observed flow values within each speed interval. The aggregation is applied to the density and speed ranges since we expect a concave flow function over each of these quantities. The aggregation process is formally described in \ref{sec:app_methodology}.

\section{Results}\label{sec:results}

The method we propose is implemented using data gathered from nine independent campaigns and across five driver modes for platooning. These clusters consist of ACC-driven vehicles with minimum, median, and maximum settings, cooperative ACC-driven vehicles, and human drivers. Although the number of observations for ACC-driven vehicles with minimum and maximum settings is significantly higher than for the other clusters, we present results incorporating all available data.  Further information regarding the data used can be found in \ref{sec:app_data} and the relevant literature.

\begin{figure*}[!ht]
    \begin{subfigure}[t]{0.45\textwidth}
        \includegraphics[width=\textwidth]{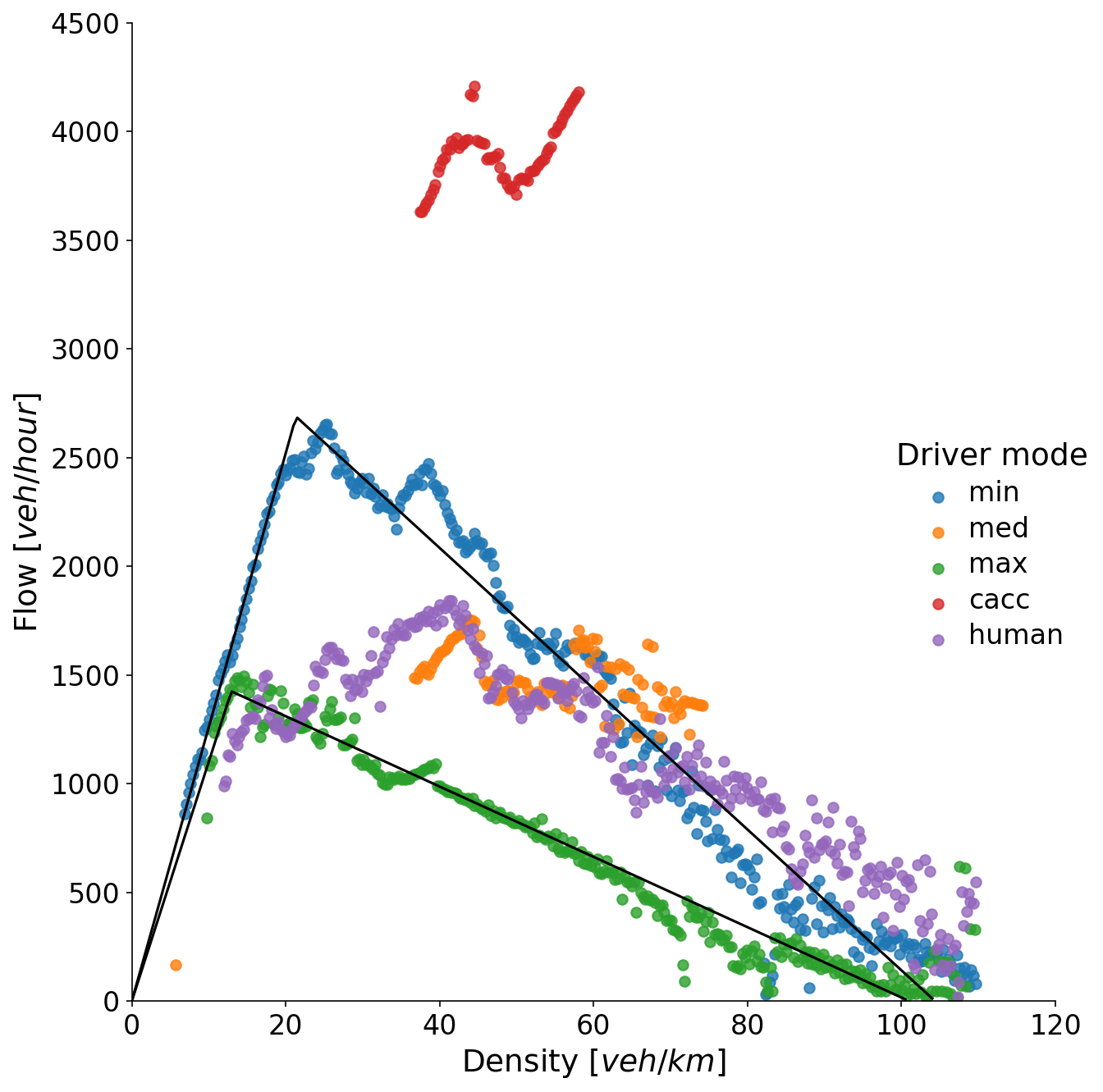}
        \caption{Proposed}\label{fig:FD_proposed}
    \end{subfigure} 
    ~ 
    \begin{subfigure}[t]{0.45\textwidth}
        \includegraphics[width=\textwidth]{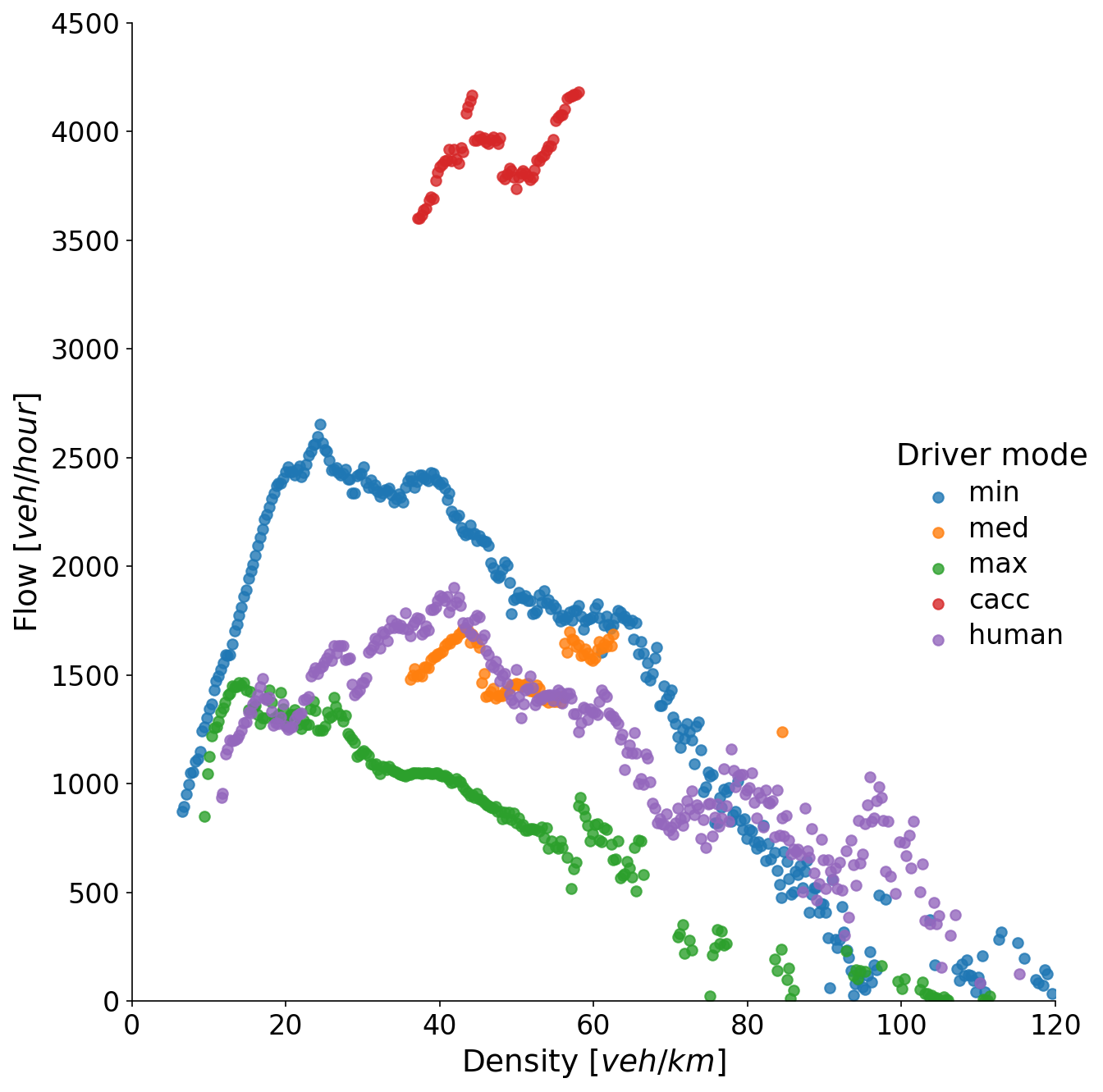}
        \caption{\cite{maiti_universality_2023}}\label{fig:FD_Laval}
    \end{subfigure}
    \caption{Flow-Density plot with the proposed method and the measurement method proposed by \cite{laval_hysteresis_2011} after the application of the Shear transformation described in \cite{maiti_universality_2023} and the aggregation process described in this paper.}
    \label{fig:results}
\end{figure*}

Figure~\ref{fig:results} illustrates the flow-density plot of the FD for two methods, the observed driver modes and a $\delta_k=\delta_v=0.3$. Please note that despite the absence of traffic state ground-truth, we compare the proposed method with \cite{laval_hysteresis_2011} as a benchmark technique and we verify the realistic traffic state inference. The inferred FDs with the proposed method (Figure~\ref{fig:FD_proposed}) and the existing state-of-the-art in the literature (Figure~\ref{fig:FD_Laval}) are very similar. The results for the other two bi-level plots, namely Speed-Density and Flow-Speed are provided in \ref{sec:app_fds}. 

The black lines in Figure~\ref{fig:FD_proposed} correspond to the triangular fundamental diagram (TFD) that was calibrated on the data from ACC-driven vehicles with a minimum and maximum setting (two driver modes with available observations). Calibration was performed with minimization of the sum of normalized root mean square errors of the flow and speed. Further information on the calibration process can be found in \ref{sec:app_calibration}.

The calibrated inferred FDs exhibit remarkable consistency with the TFD shape. However, depending on the aggregation strength (defined by $\delta$), some TFD parameters might lose their significance during calibration when there is insufficient available data. For instance, the observed speeds for the maximum ACC setting only range up to 120~$[km/h]$, while for the minimum ACC setting, they reach 140~$[km/h]$, see Figure~\ref{fig:FD_vk_proposed}~and~\ref{fig:FD_vq_proposed}. As a result, the calibration of the free-flow part for the former might not be accurate. Similarly, it is currently not meaningful to provide calibrated TFD for the other driver modes, namely, medium ACC setting and CACC, due to scarcity of data.

Another interesting observation relates to the calibrated backward wave speed values for the ACC-driven vehicles that are given in Table~\ref{tab:sensitivity}. ACC-driven vehicles with maximum setting have realistic wave speed of around 15-17$km/h$. However, the same variable for ACC-driven vehicles with minimum setting has been found to be over 30$km/h$. This reveals safety concerns for such platoons. An in-depth investigation is out of the present paper's scope but will be part of future research. 

While the available data for human drivers may seem sufficient, it appears that the TFD shape only poorly describes their behavior. The respective FD has a more smooth shape with a very wide near-capacity area. The behavioral analysis of the FD for human drivers reveals commonalities and differences between human behavior and control behavior which eventually is also reflected in traffic flow. By referring to Figure~\ref{fig:app_FDs} (provided in \ref{sec:app_fds}), it can be observed that human drivers behave similarly to ACC-driven vehicles with a maximum setting at higher speeds (above 50~$[km/h]$). However at lower speeds (below 30~$[km/h]$), reflecting congested conditions, human drivers take greater safety risks by maintaining much shorter distances from vehicles ahead. Consequently, their behavior aligns more closely with ACC-driven vehicles with a minimum setting. In terms of traffic implications, such behavior leads to better road utilization compared to partially automated vehicles under over-saturated conditions nearing jam density. Unfortunately, there is no available data to observe the corresponding behavior of CACC-driven vehicles but it would be interesting to explore this in future research. Finally, the above analysis conceptually aligns with empirical evidence, showing human drivers being more relaxed at higher speeds, allowing for larger time gaps, while being more agile at lower speeds to minimize their travel time, see \cite{xu_analysis_2021}.

Consistent with existing literature, the findings presented in this study suggest that connectivity in ACC systems has the potential to improve traffic flow. Although the empirical data available is limited, it is evident that the observed behavior of CACC-driven vehicles differs significantly from both ACC-driven and human-driven vehicles. While we cannot derive precise values for capacity, critical density, and jam density due to the scarcity of data, it is evident that the potential for improvement is substantial. The observed flow reaches around 4000~$[veh/h]$ (possibly higher overall) as can be seen in Figure~\ref{fig:results}. Furthermore, Figure~\ref{fig:app_FDs} in Annex~\ref{sec:app_fds} shows that they manage to maintain the same platoon speed as the other modes (ACC and human drivers) under very high density, thus very close inter-vehicle spacing. This, of course, can be attributed to the effect of connectivity that provides more information to the vehicles, so that, they can keep a close distance from their leaders without compromising safety and stability. It should be noted, however, that further empirical data is necessary to validate the safety and robustness of CACC systems in the face of communication interference threats.

It is important to acknowledge that the nominal capacity values mentioned earlier assume ideal uniform driving conditions, which are not reflective of real-world environments. However, the relevant FD characteristics provide valuable insights into the variations in efficiency among different driver modes of transportation and the potential each mode holds. Such information is essential for the design and implementation of optimal traffic management and control strategies.

\subsection{Analysis of different platoon states}\label{sec:platoon_states}

In the literature, the estimation of stationary conditions in a platoon dataset is necessary for the inference of traffic variables as mentioned in \cite{laval_hysteresis_2011}. However, this is not a requirement for the proposed method. Nevertheless, during platoon acceleration, deceleration, and stable (cruising) states, different behaviors arise and therefore it is interesting to be able to observe if and how such states affect a unified FD representation. Consequently, we post-process all available data and categorize the platoon trajectories into three platoon states, namely acceleration, deceleration, and stable state events. Categorization is performed by inferring partial trajectories where the average platoon speed is increasing, decreasing, or remaining constant. The idea is then to aggregate traffic variables per platoon state so that we analyze their impact on the unified FD representation. For more in-depth information about the processing details, we refer the reader to Annex~\ref{sec:app_states}.

\begin{figure*}[h]
    \begin{subfigure}[t]{0.65\textwidth}
        \includegraphics[width=\textwidth]{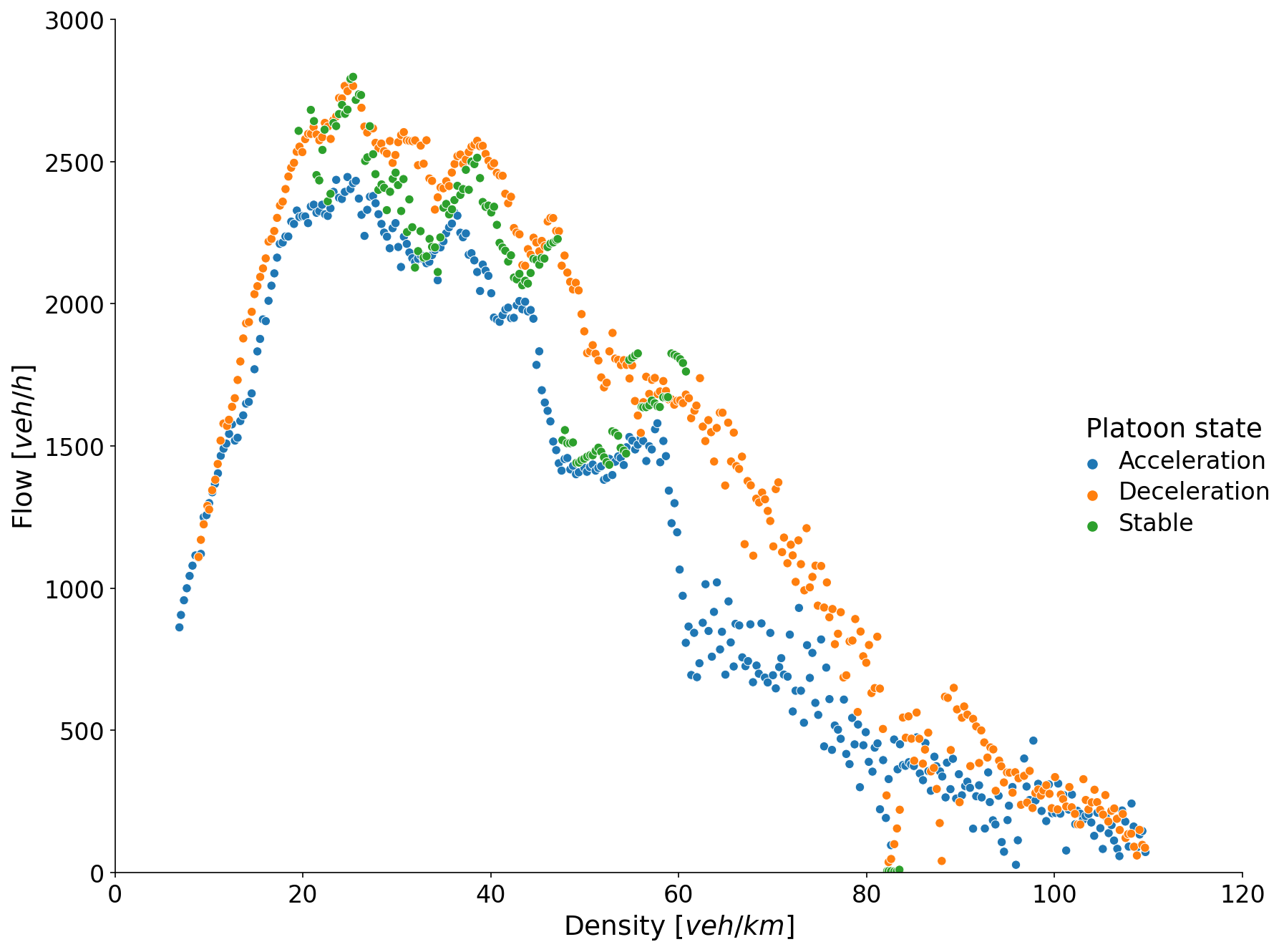}
        \caption{Flow-Density plot - proposed}
    \end{subfigure}
    \caption{Flow-Density plot for different platoon states. The plots refer to observed vehicle platoons driven with the ACC minimum setting.}
    \label{fig:platoon_states_FD_QK}
\end{figure*}

The results with the proposed method per platoon state for ACC-driven vehicles with minimum setting are illustrated in Figure~\ref{fig:platoon_states_FD_QK}. The traffic states corresponding to stable platoon state lie between those of accelerating and decelerating states, at (visually) equal distances. Interestingly, by looking only at stable platoon states, we would not have been able to infer the complete FD.

\subsection{\texorpdfstring{Sensitivity of $\delta$ parameter in calibrated FD}{Sensitivity of delta parameter in calibrated FD}}\label{ssec:sensitivity}

The results presented so far refer to $\delta_k$ and $\delta_v$ parameters equal to 0.3~$[veh/km]$ and 0.3~$[km/h]$ values for data aggregation. Here we test the consistency of the calibrated TFD parameters through a sensitivity analysis for different $\delta$ values ranging from 0.3 to 3.5 density/speed units. The results validate completely the consistency of the inferred FD in all cases and they are summarized in Table~\ref{tab:sensitivity} in \ref{sec:app_sensitivity}. It is interesting to observe that the calibrated TFD parameters for the minimum and maximum $\delta$s are really close, even if higher $\delta$ leads to very few points for calibration.

\section{Conclusion}\label{sec:conclusion}

We study different experimental observations with platoons of vehicles driven by five driver modes, i.e., Adaptive Cruise Control (ACC) with minimum, medium and maximum setting, Cooperative-ACC (CACC) and human drivers. We propose a numerically simple and effective method to derive the Fundamental Diagram (FD) per driver mode after the application of a bivariate aggregation technique. The
proposed method demonstrates a remarkably consistent relation between the traffic variables and
a clear triangular shape for autonomously-driven vehicles. Comparison with a state-of-the-art method from the literature confirms the validity of the inferred states. Two important benefits are the absence of a need to assume a backward wave speed and to estimate stationary conditions.

The driver mode with minimum ACC setting points to much higher capacity values (roughly 90\%) than ACC-driven vehicles with maximum setting or human drivers. However, this also leads to much higher backward wave speed, raising safety concerns. Preliminary results with limited data show that connectivity can significantly improve capacity. The FD shape for automated vehicles is triangular, while human drivers exhibit a wider near-capacity area. 

\clearpage
\appendix

\section{Methodology}
\label{sec:app_methodology}
Edie's generalized definition is given for reference below:

\begin{subequations}
\begin{align}
k &= \sum_{i=1}^{N} t_i\Big/\mid A\mid \label{eq:Edie_k}\\
q &= \sum_{i=1}^{N} x_i\Big/\mid A\mid \label{eq:Edie_q}\\
v &= q/k = \sum_{i=1}^{N} x_i\Big/\sum_{i=1}^{N} t_i  \label{eq:Edie_v}
\end{align}
\end{subequations}

\noindent where $k$, $q$ and $v$ stand for density, flow, and speed in the region: $A$, $\mid A\mid$ is the area of $A$ and $t_i, x_i$ is the travel time and distance traveled respectively, of vehicle $i$ inside $\mid A\mid$. 

The effective platoon length $l_p(t)$, is shown in Figure~\ref{fig:area} and it is dynamic over time as the vehicles move. We define $\mid A\mid$ as a time-variant scalene trapezoid area described but two consecutive platoon steps. Such a simplification is by design particularly useful and inexpensive for the analysis of discrete trajectory observations.

Let us assume that we have platoon observations with $N$ vehicles and a constant data acquisition frequency equal to $f=\frac{1}{\delta t}$. Figure~\ref{fig:area} provides a schematic representation of the area used for the computation of speed, density, and flow. Considering that the platoon $l_p(t)$ at two consecutive time instances, we derive $\mid A\mid$ numerically as follows:

\begin{equation}\label{eq:area}
\begin{gathered}
    \mid A(t)\mid = \frac{l_p(t)+l_p(t + \delta t)}{2} \delta t 
\end{gathered}
\end{equation}

Speed, density, and flow variables for a platoon of $N$ vehicles and for a fixed observation interval $\delta t$ are computed accordingly:

\begin{subequations}
\begin{align}
k(t) &= (N \delta t) \Big/\mid A(t)\mid = 2N \Big/ (l_p(t)+l_p(t + \delta t)) \label{eq:proposed_k}\\
q(t) &= \sum_{i=1}^{N} \delta x_i(t)\Big/\mid A(t)\mid \label{eq:proposed_q}\\
v(t) &= q(t)/k(t) = \sum_{i=1}^{N} \delta x_i(t)\Big/ (N \delta t) \label{eq:proposed_v}
\end{align}
\end{subequations}

where $\delta x_i(t) = x_i(t+\delta t)-x_i(t)$, derived by the individual vehicle positions at times $t$ and $t + \delta t$. 

In the remaining part of this Appendix, we describe formally the aggregation process applied on the available observation per driver mode. Let us define a set of $M + 1$ discrete equally spaced values as follows: 

\begin{equation}\label{eq:bins}
\begin{gathered}
    B = \{0*\delta, 1*\delta, \ldots M*\delta\} 
\end{gathered}
\end{equation}

where $\delta$ is the discrete quantity interval and $B$ is the corresponding set of discrete quantity values. In our case, we create two sets, one for density discretization and another for speed discretization, with the corresponding $\delta_k$ and $\delta_v$ intervals, i.e. $B_k = \{0*\delta_k, 1*\delta_k, \ldots M*\delta_k\}$ and $B_v = \{0*\delta_v, 1*\delta_v, \ldots M*\delta_v\}$.

In order to compute the aggregated values for $v$ - $k$ and $q$ - $k$ plots based on observed density, speed and flow variables, i.e. $k(t)$, $v(t)$, $q(t)$, for each $\delta_k$ within $B_k$, we average all available measurements across all datasets to derive $M+1$ points that describe the central data tendency.

\begin{equation}\label{eq:bin_agg1}
  \{k,v,q\}_i = 
   \begin{cases} 
   \frac{1}{T_i}\sum_{t=1}^{T_i}k(t), & \mbox{where } i*\delta_k \leq k(t) \leq (i+1)*\delta_k, \forall i \in \{0, \ldots M_k\}  \\
   \frac{1}{T_i}\sum_{t=1}^{T_i}v(t), & \mbox{where } i*\delta_k \leq k(t) \leq (i+1)*\delta_k, \forall i \in \{0, \ldots M_k\}  \\
   \frac{1}{T_i}\sum_{t=1}^{T_i}q(t), & \mbox{where } i*\delta_k \leq k(t) \leq (i+1)*\delta_k, \forall i \in \{0, \ldots M_k\}  \\
   \end{cases}
\end{equation}

where $T_i$ refers to the total number of $k(t)$, $v(t)$, $q(t)$ measurements when $k(t)$ lies within $(i*\delta_k,(i+1)*\delta_k]$.

Finally, for the computation of aggregated values for $v$ - $q$, since it is expected a two-valued function across the flow, the aggregation is performed on the speed in a similar way as Eq.~\ref{eq:bin_agg1} with $\delta_v$ interval according to the following equation:

\begin{equation}\label{eq:bin_agg2}
  \{v,q\}_i = 
   \begin{cases}  
   \frac{1}{T_i}\sum_{t=1}^{T_i}v(t), & \mbox{where } i*\delta_v \leq v(t) \leq (i+1)*\delta_v, \forall i \in \{0, \ldots M_v\}  \\
   \frac{1}{T_i}\sum_{t=1}^{T_i}q(t), & \mbox{where } i*\delta_v \leq v(t) \leq (i+1)*\delta_v, \forall i \in \{0, \ldots M_v\}  \\
   \end{cases}
\end{equation}

where $T_i$ refers to the total number of available $k(t)$, $v(t)$, $q(t)$ measurements when $v(t)$ lies within $(i*\delta_v,(i+1)*\delta_v]$. Please note that $M_k=M_v$ in the case that $\delta_k=\delta_v$. For simplicity, in the results of this work, we use the same $\delta$ values on the density and flow axes.

\section{Data}\label{sec:app_data}

The proposed methodology is demonstrated through empirical observations from car-following campaigns with platoons of vehicles. Eleven campaigns from five major datasets are considered in this work. The observations refer to human-driven, ACC-driven and CACC-driven vehicles. The total distance traveled per platoon in all experiments is 844.8~$[km]$ for ACC-driven platoons, 38.2~$[km]$ for human-driven platoons and 29.3~$[km]$ for CACC-driven platoons. A brief introduction of each campaign follows while all datasets are summarized in TABLE~\ref{tab:campaigns}. 

\begin{table}[!ht]
	\caption{Summary of experimental campaigns used in this work.}\label{tab:campaigns}
	\smaller
	\begin{center}
		\setlength{\arrayrulewidth}{.15em}
        \def\arraystretch{1.7}
		\begin{tabular}{l | c c c c c c }
    		 & Driver & ACC settings & Number of vehicles & Homogenous\\\hline
		 \textit{AstaZero} & ACC:172~$[km]$/HV:22.9~$[km]$ & min & 5 & No\\
		 \textit{ZalaZone} (Handling) & ACC:49.5~$[km]$/HV:15.3~$[km]$ & min/max & 4-10 & No\\ 
		 \textit{ZalaZone} (Dynamic) & ACC:67.3~$[km]$ & min/med/max & 4-11 & No\\ 
		 \textit{Cherasco} & ACC:6.5~$[km]$ & min & 3 & No\\ 
		 \textit{Casale} & ACC:142.5~$[km]$ & min & 2 & Yes\\ 
		 \textit{JRC} & ACC:10.9~$[km]$ & min/med & 2/3 & Yes/No\\ 
		 \textit{CARMA} & CACC:29.3~$[km]$ & - & 5 & Yes\\ 
		 \textit{VU1} & ACC:390~$[km]$ & min/max & 2 & Yes\\ 
		 \textit{VU2} & ACC:6.1~$[km]$ & min & 7 & Yes\\
		\end{tabular}
	\end{center}
\end{table} 

\subsubsection{OpenACC dataset}

The OpenACC dataset was initiated by the open-access policy of the Joint Research Centre of the European Commission and involves a set of car-following experiments with human-driven and ACC-driven vehicles on public roads, test tracks, and the JRC private urban network. The dataset is constantly populated and this work includes campaigns that are not mentioned in the original publication. In this work, we use car-following observations from AstaZero, ZalaZone, Cherasco, Casale, and JRC campaigns. More details about the OpenACC can be found in \cite{makridis_openacc_2021}. Currently, available campaigns can be downloaded from the \cite{makridis_open_2020} website.

\subsubsection{CARMA dataset}
The United States Department of Transportation has developed a proof-of-concept platooning system that, additionally to the familiar ACC, includes V2V and V2I communication. This CACC implementation is based on the Cooperative Automation Research and Mobility Applications (CARMA) platform. As described in \citet{tiernan_test_2017}, the first version of \textit{CARMA} dataset was acquired in 2016 at the Aberdeen Test Center in Maryland. In this work, we use the second subsequent campaign, CARMA2, a platooning experiment presented in \citet{tiernan_carma_2019} in 2018. The experiment included five vehicles of the same model and was conducted on the 7.2~$[km]$ two-lane road of the proving ground. The features of the test track are assumed to be similar to the typical US highway. Spacing measurements were obtained with PinPoint GPS, while for speed measurements the production wheel speed was used. More details on the dataset can be found in the above-mentioned reports.  

\subsubsection{Vanderbilt University dataset}
The \textit{Vanderbilt University dataset} includes two campaigns with car-following data of ACC-driven vehicles. During the first campaign (VU1) four speed profiles are recorded to observe the behavior of each vehicle in the two-vehicle tests. For all tests, the vehicles begin on the track and start at a low speed with a full-size sedan as the lead vehicle, and the test vehicle as the follower vehicle. To validate the emergent traffic flow behavior of ACC vehicles at the aggregate (system) level, a second campaign (VU2) was conducted where seven identical ACC vehicles follow a lead vehicle that creates a velocity slow down event. Position and speed data for each vehicle were collected using high-accuracy uBlox EVK-M8T GPS receivers. For more information we refer the reader to the corresponding publication \cite{gunter_are_2021}.



\section{Fundamental Diagrams}\label{sec:app_fds}

\begin{figure*}[h]
    \begin{subfigure}[t]{0.45\textwidth}
        \includegraphics[width=\textwidth]{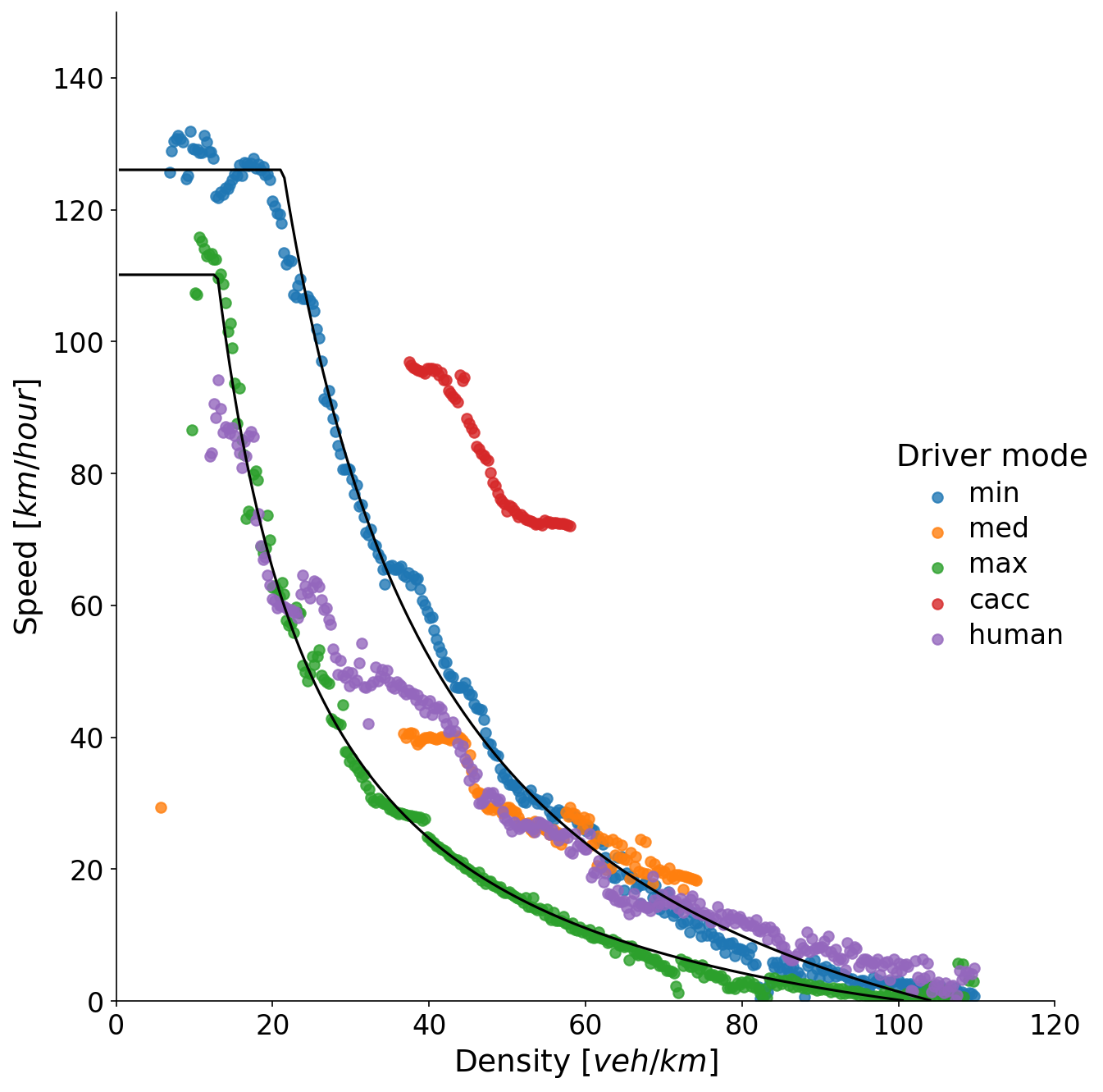}
        \caption{ $v$ - $k$ plot - proposed}\label{fig:FD_vk_proposed}
    \end{subfigure}
    ~ 
    \begin{subfigure}[t]{0.45\textwidth}
        \includegraphics[width=\textwidth]{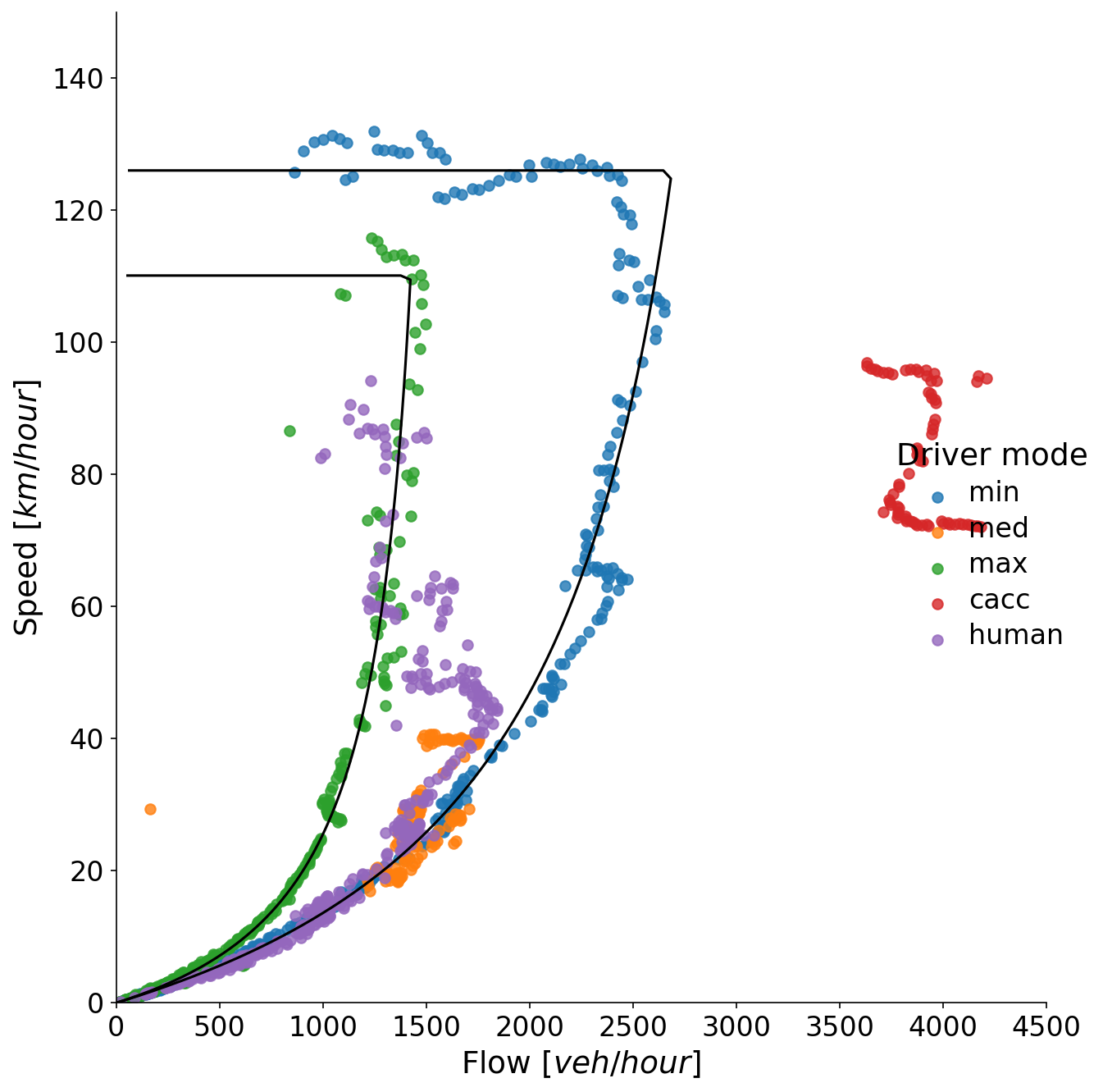}
        \caption{ $v$ - $q$ plot - proposed}\label{fig:FD_vq_proposed}
    \end{subfigure}
    ~
    \begin{subfigure}[t]{0.45\textwidth}
        \includegraphics[width=\textwidth]{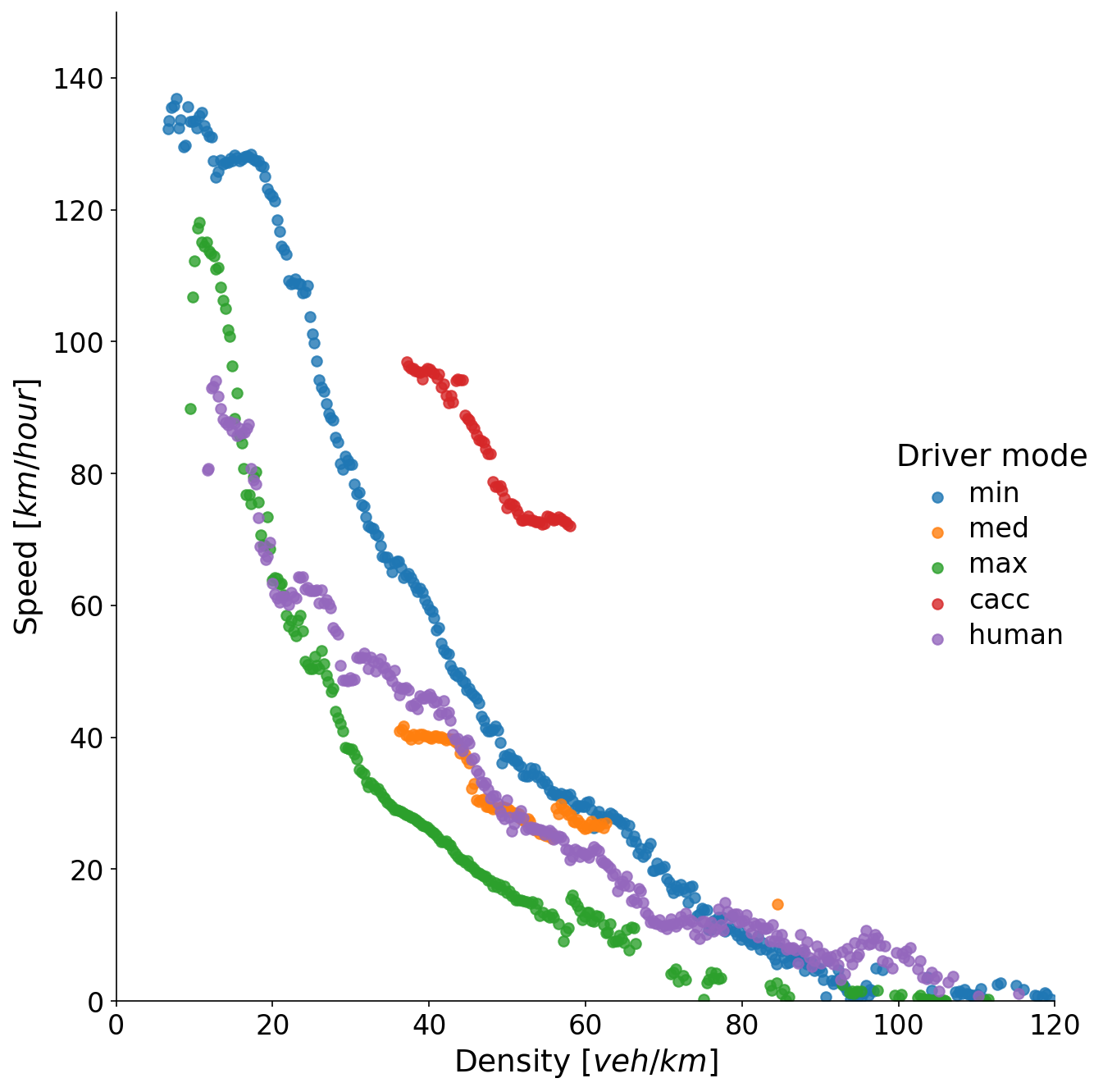}
        \caption{$v$ - $k$ plot - \cite{maiti_universality_2023}}\label{fig:FD_vk_Laval}
    \end{subfigure}
    ~ 
    \begin{subfigure}[t]{0.45\textwidth}
        \includegraphics[width=\textwidth]{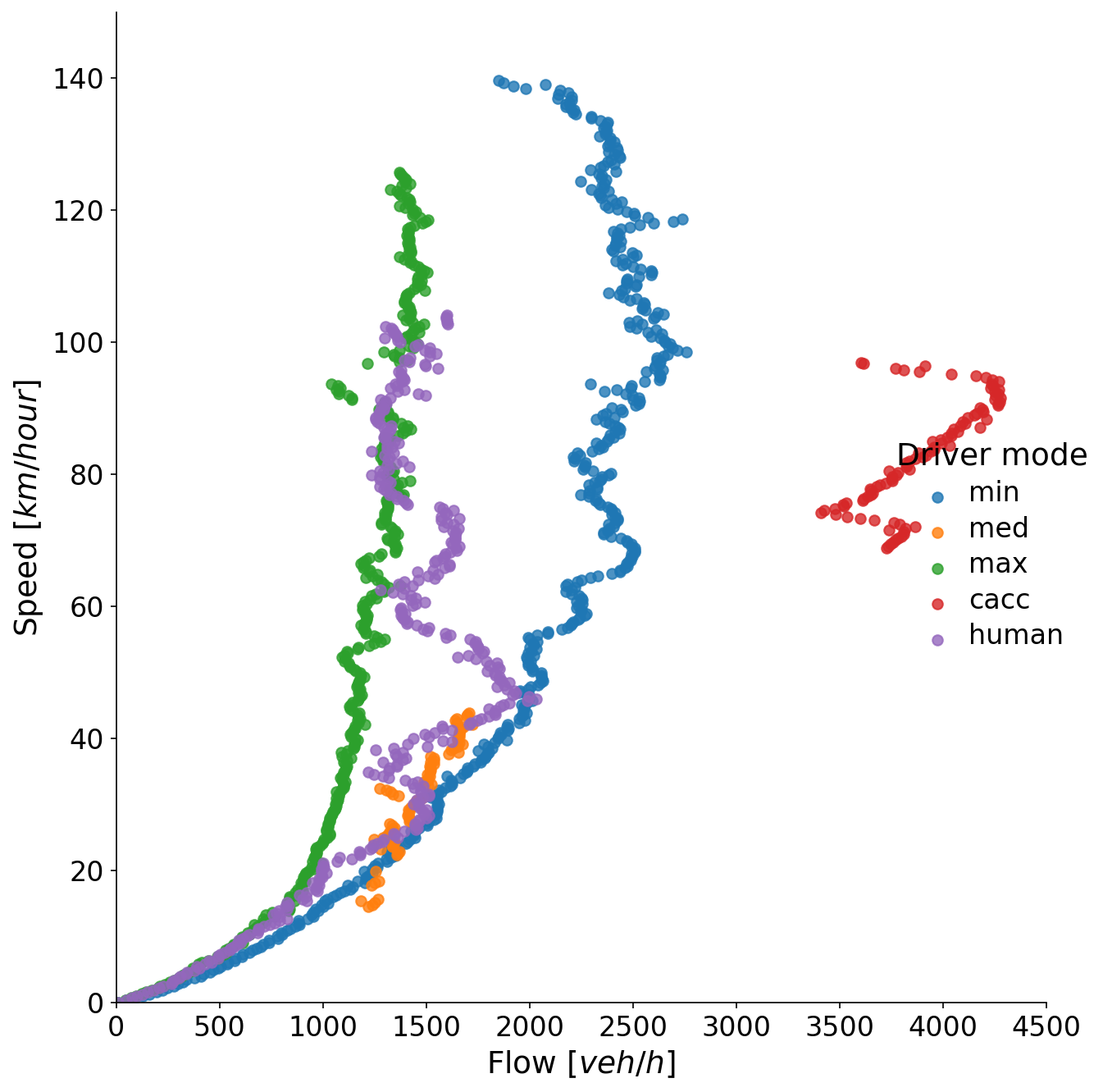}
        \caption{$v$ - $q$ plot - \cite{maiti_universality_2023}}\label{fig:FD_vq_Laval}
    \end{subfigure}
    \caption{Bi-level plots with the proposed method after aggregation}.
    \label{fig:app_FDs}
\end{figure*}

The flow-density plots are given in Figure~\ref{fig:FD_proposed}~and~\ref{fig:FD_Laval}. Here we provide additionally the Speed-Density and Flow-Speed plots for both methods in Figure~\ref{fig:app_FDs}. Calibrated FDs fit very nicely on the data with a minimum and maximum ACC setting. Unfortunately, there are limited data on CACC-driven vehicles but figures show great potential for increased capacity and critical density for this technology. It can safely be assumed that this result is due to the additional information from vehicle connectivity As mentioned in Section~\ref{sec:results}, the FD for human 
drivers seems to be between the FD of ACC-driven vehicles with minimum setting and those with a maximum setting. Attempting to interpret this finding intuitively, we can assume that humans drive keeping larger time spaces at high speeds, while they keep shorter spacings at low speeds, i.e., the congested branch of the FD. Finally, as expected ACC-driven vehicles with a medium setting generate FD points between those with a maximum and a minimum setting.

\section{Calibration of the Triangular Fundamental Diagram}\label{sec:app_calibration}

For calibration purposes, we use the triangular fundamental diagram (TFD) proposed by \cite{daganzo_requiem_1995}, which is probably the most widespread. The TFD has been chosen due to its simplicity in the implementation, describing the density-flow plane with a bi-linear relation. More specifically, the flow-density function is adequately described by the following:

\begin{equation}\label{eq:TFDq}
  Q(k) = 
   \begin{cases}  
   v_f k, & \mbox{if } k \leq k_{cr}  \\
   w (k_{jam} -k), & \mbox{if } k \geq k_{cr}  \\
   \end{cases}
\end{equation}

where $v_f$ is the free-flow speed, $w$ is the wave speed, $k_{cr}$ is the critical density and $k_{jam}$ is the jam density. 

Based on trigonometry and assuming having critical density $k_{cr}$, the first and second leg of Equation~\ref{eq:TFDq} are equal. Therefore, we can derive the following equation for the wave speed and reduce the number of variables by one:

\begin{equation}\label{eq:TFDw}
\begin{gathered}
    w = \frac{v_f k_{cr}}{k_{jam} - k_{cr}}
\end{gathered}
\end{equation}

From Equations~\ref{eq:TFDq}~and~\ref{eq:TFDw} we have:

\begin{equation}\label{eq:TFDq2}
  Q(k) = 
   \begin{cases}  
   v_f k, & \mbox{if } k \leq k_{cr}  \\
   \frac{v_f k_{cr}}{k_{jam} - k_{cr}}k, & \mbox{if } k \geq k_{cr}  \\
   \end{cases}
\end{equation}

Consequently we perform the following simple optimization operation on the $v$-$k$ (see Figure~\ref{fig:results}) and the parameter vector $x = [v_f, k_{cr}, k_{jam}]$, which contains the parameters of the TFD:

\begin{equation}\label{eq:optmin}
x = \argmin_{x: \text{lb} \leq x \leq \text{ub}} \mathcal{Q_{x}}
\end{equation} 

where $\text{lb}$ and $\text{ub}$ are the lower and upper bounds for the three variables of the TFD. $\mathcal{Q_{x}}$ is the objective function, based on the root mean square errors between observations and the TFD:

\begin{equation}\label{eq:vq}
\mathcal{Q_{x}} = \frac{\sqrt{\frac{1}{(M+1)}\sum_{m=0}^{M}(q_m-Q(k_m))}}{\frac{1}{(M+1)}\sum_{m=0}^{M}(q_m)} +
\frac{\sqrt{\frac{1}{(M+1)}\sum_{m=0}^{M}(v_m-Q(k_m)/k_m)}}{\frac{1}{(M+1)}\sum_{m=0}^{M}(v_m)} 
\end{equation} 

\section{Analysis of acceleration, deceleration and stable platoon states}\label{sec:app_states}

\begin{figure*}[!b]
    \begin{subfigure}[t]{0.45\textwidth}
        \includegraphics[width=\textwidth]{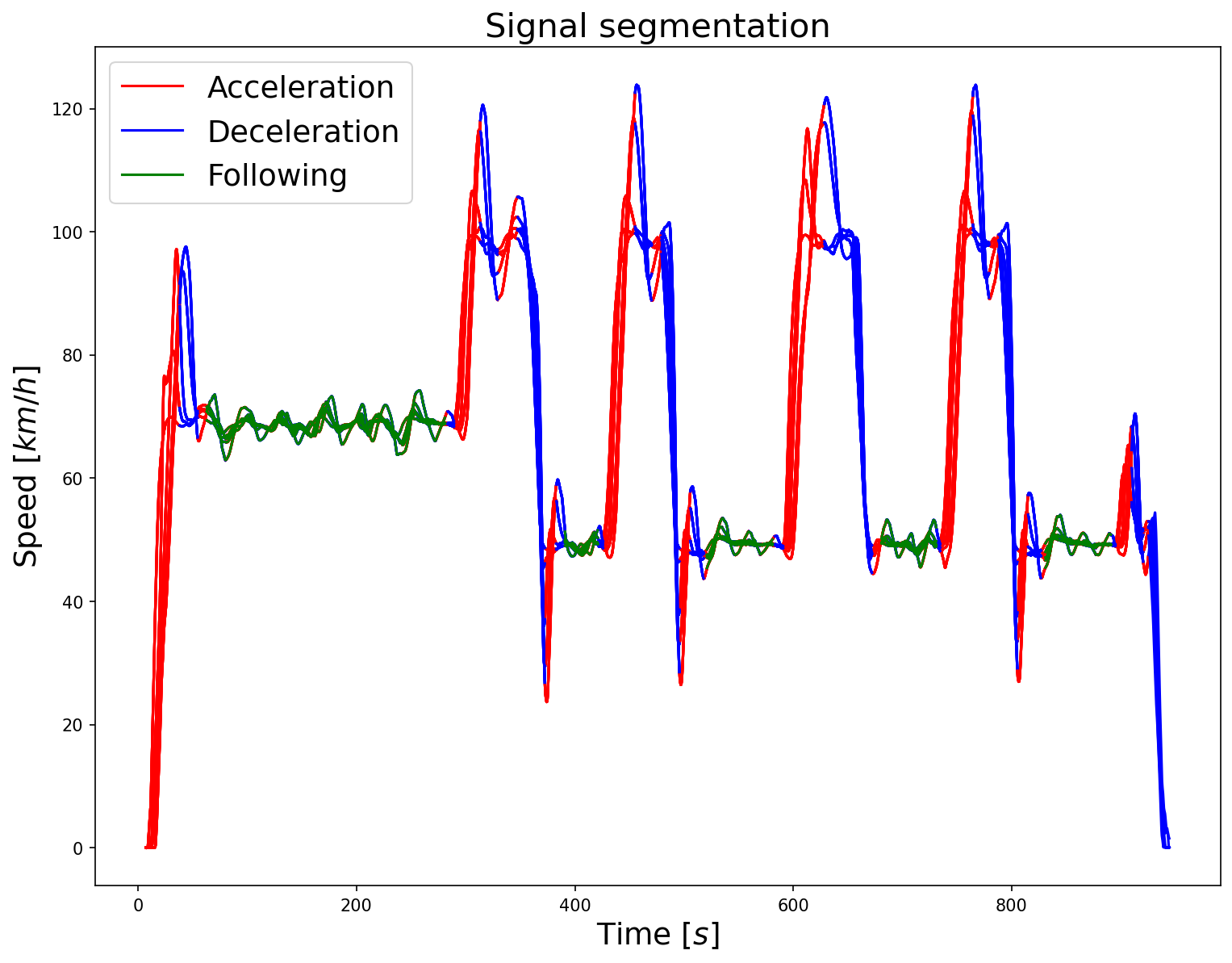}
        \caption{Speed profile segmentation based on platoon average speed in acceleration, deceleration and following state events.}\label{fig:split}
    \end{subfigure}
    ~ 
    \begin{subfigure}[t]{0.45\textwidth}
        \includegraphics[width=\textwidth]{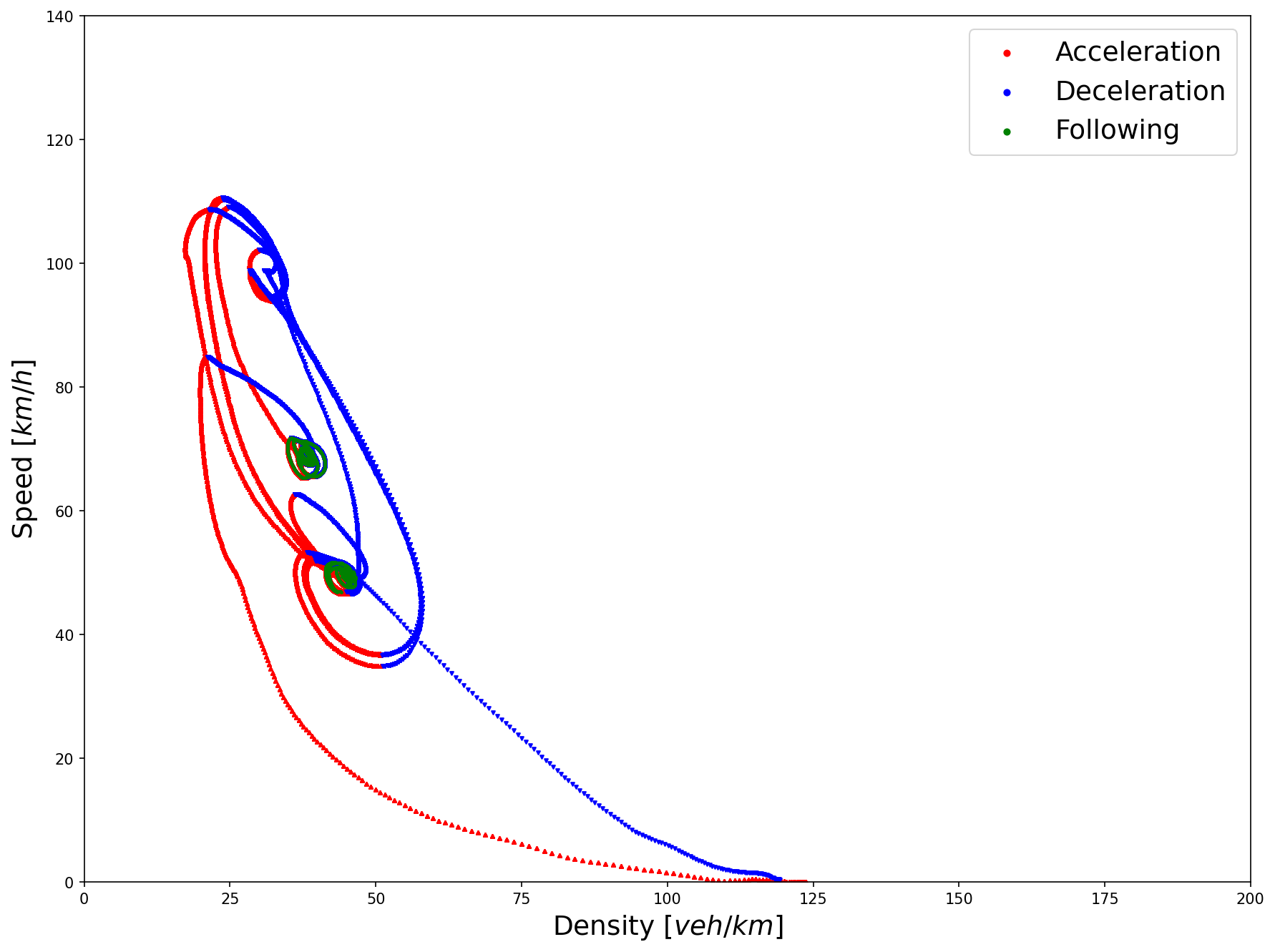}
        \caption{The derived $v$ - $k$ diagram for a partial platoon trajectory from the \textit{AstaZero} campaign.}\label{fig:vk_AstaZero}
    \end{subfigure}
    \caption{Acceleration, deceleration and stable platoon states.}
    \label{fig:platoon_states}
\end{figure*}

Figure~\ref{fig:split} shows the segmented speed profiles from a five-vehicle platoon from the \textit{AstaZero} campaign. The segmentation was performed on the basis of the average platoon speed (platoon profile). We obtain acceleration and deceleration parts using a local min/max algorithm on the platoon profile~\footnote{https://www.csc.kth.se/~weinkauf/notes/persistence1d.html}. Furthermore, we manually label the stable parts, that is, partial speed profiles where the vehicles follow each other in platoon formation with only slight speed variations. 

\begin{figure*}[!b]
    \begin{subfigure}[t]{0.45\textwidth}
        \includegraphics[width=\textwidth]{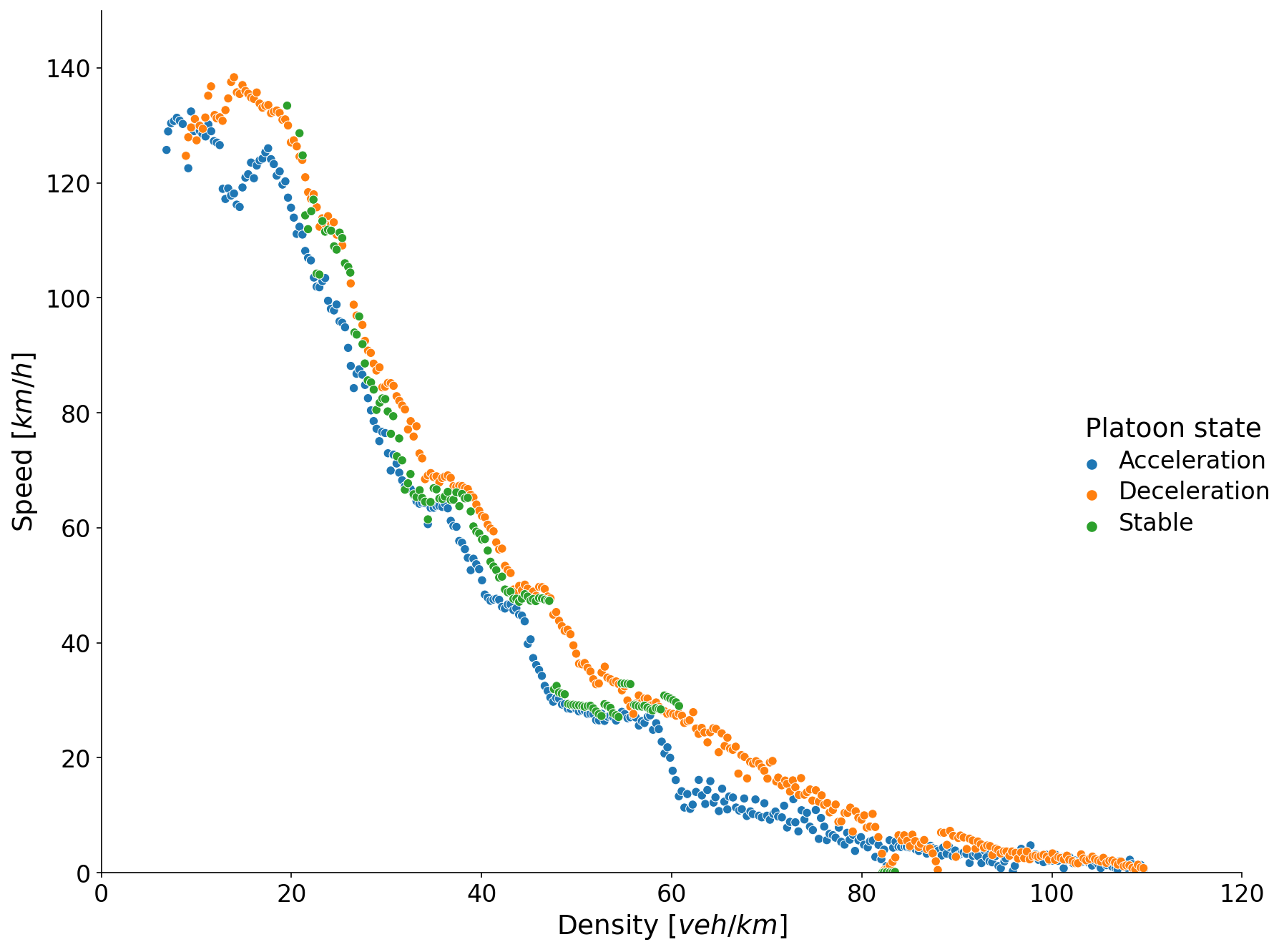}
        \caption{$v$ - $k$ plot - proposed}\label{fig:states_speed_density}
    \end{subfigure}
    ~ 
    \begin{subfigure}[t]{0.45\textwidth}
        \includegraphics[width=\textwidth]{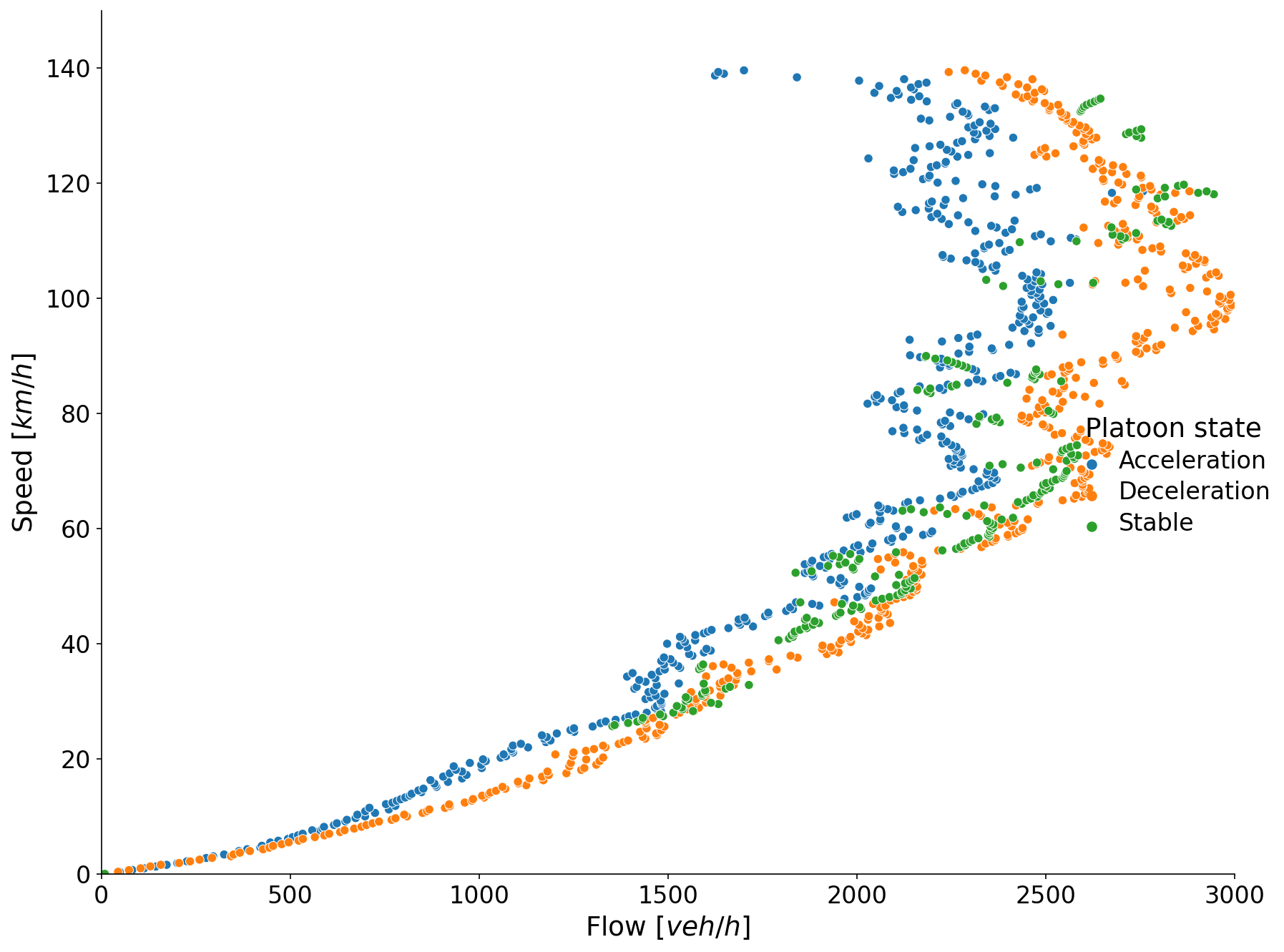}
        \caption{$q$ - $v$ plot - proposed}\label{fig:states_flow_speed}
    \end{subfigure}
    \caption{Bi-level plots for different platoon states. The plots refer to observed vehicle platoons driven with the ACC minimum setting.}
    \label{fig:platoon_states_bilevel}
\end{figure*}

Figure~\ref{fig:vk_AstaZero} shows the speed-density ($v$-$k$) plot as they are computed by Eq.~\ref{eq:proposed_k} and Eq.~\ref{eq:proposed_v} respectively, for the platoon measurements shown in Figure~\ref{fig:split}. The colors correspond to the three different states mentioned above. Noticing the green points that correspond to equilibrium conditions, it is already visible a quasi-linear relation between platoon production and platoon density, (similar to the speed-spacing relation in the analysis of trajectories at link level). The speed-density and flow-speed plots are shown for completion in Figure~\ref{fig:platoon_states_bilevel}.

\section{Sensitivity analysis on the aggregation process}\label{sec:app_sensitivity}

\begin{figure*}[t]
    \begin{subfigure}[t]{0.32\textwidth}
        \includegraphics[width=\textwidth]{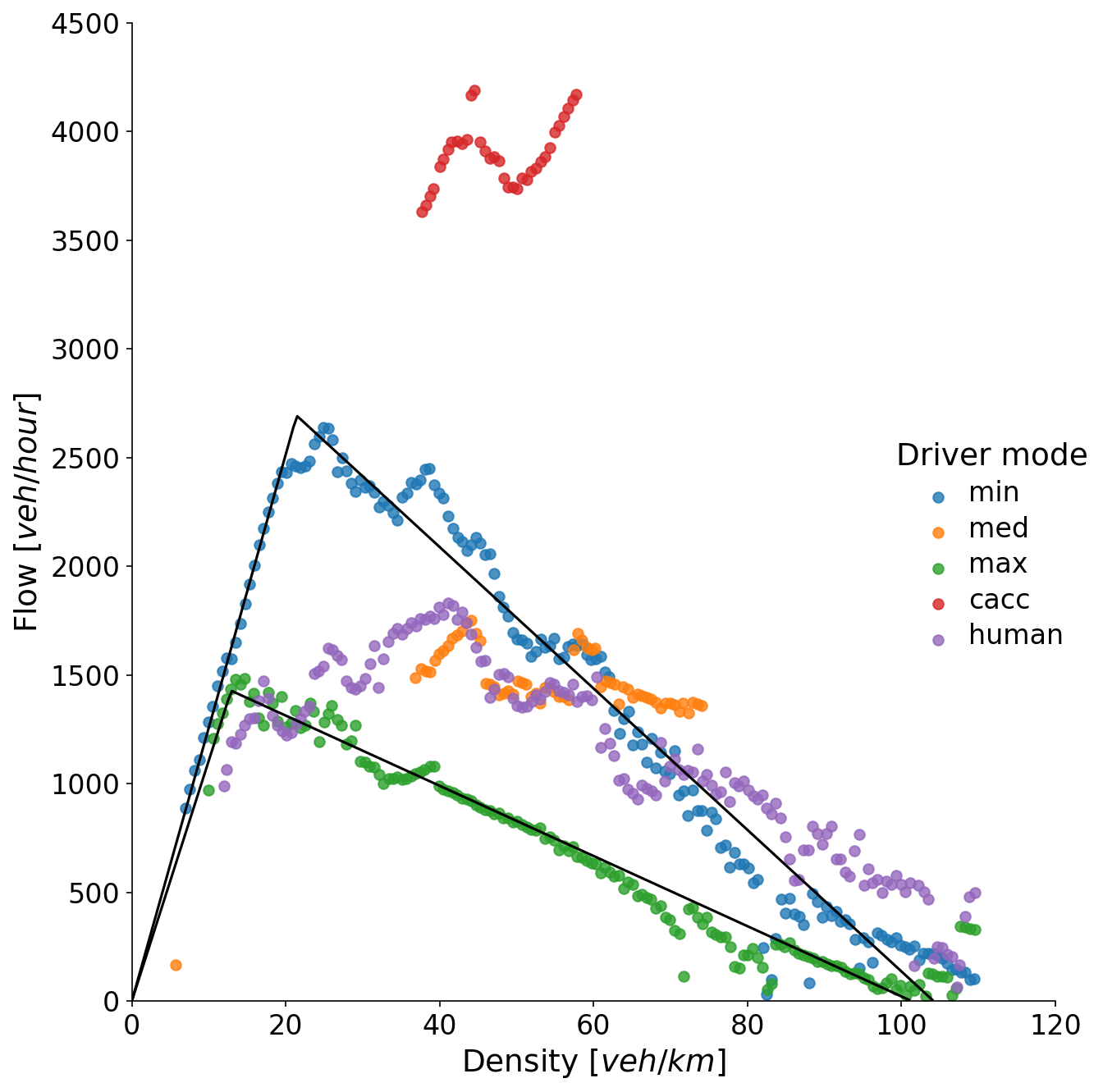}
        \caption{$q$ - $k$ plot with $\delta_{k,v}=0.6$.}\label{fig:vk_delta06}
    \end{subfigure}
    ~ 
    \begin{subfigure}[t]{0.32\textwidth}
        \includegraphics[width=\textwidth]{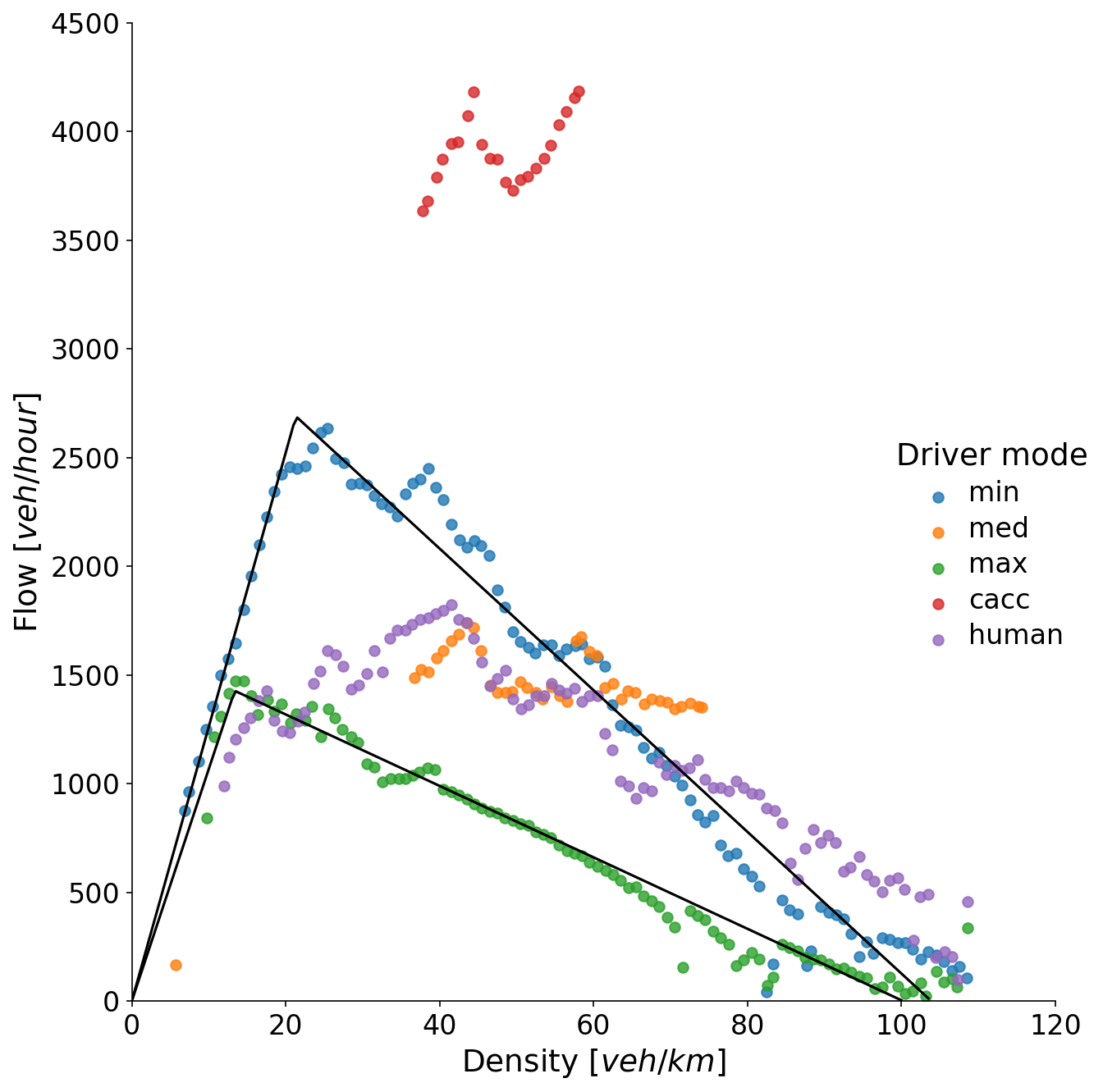}
        \caption{$q$ - $k$ plot with $\delta_{k,v}=1.0$.}\label{fig:vk_delta09}
    \end{subfigure}
     ~ 
    \begin{subfigure}[t]{0.32\textwidth}
        \includegraphics[width=\textwidth]{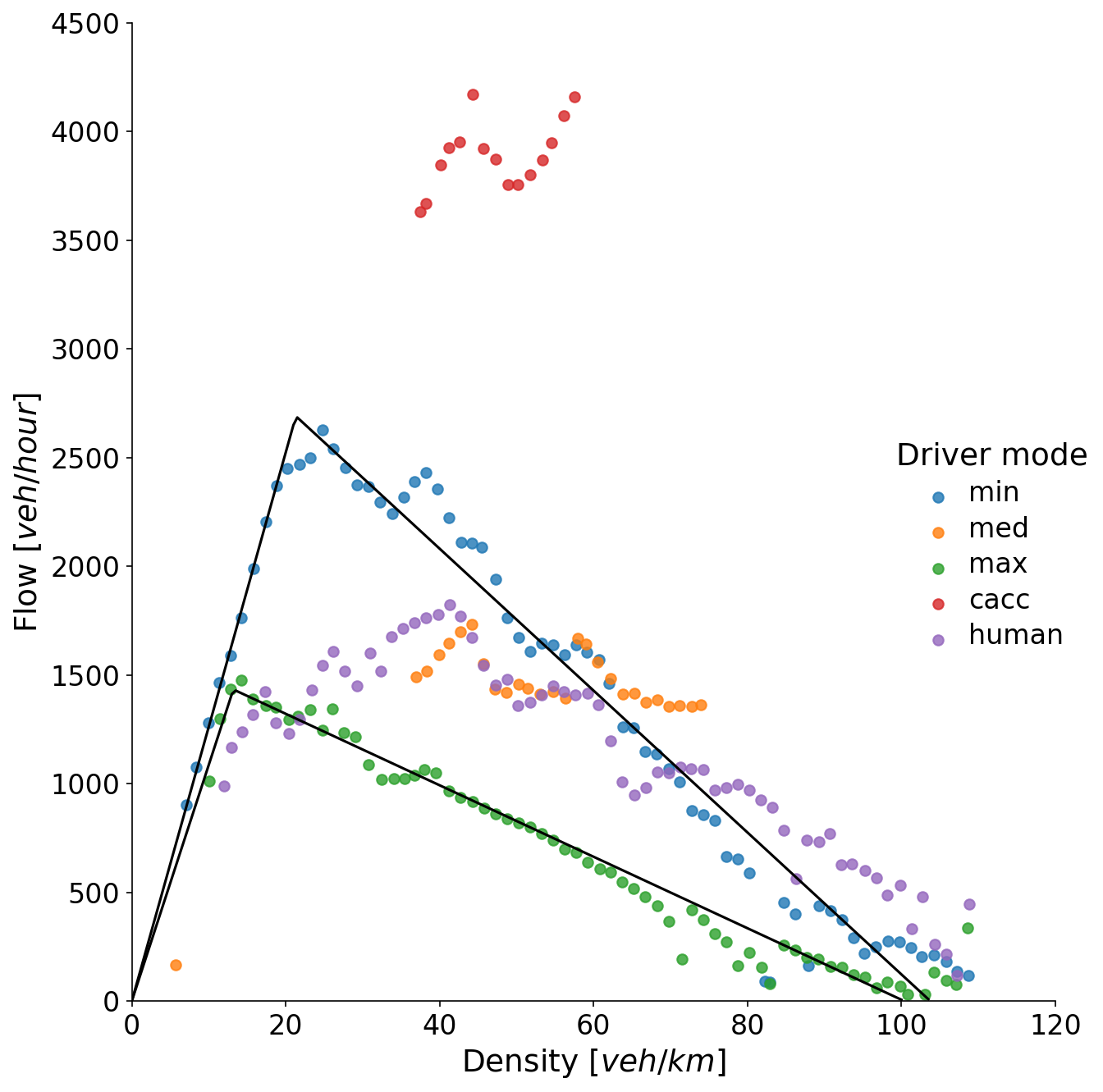}
        \caption{$q$ - $k$ plot with $\delta_{k,v}=1.5$.}\label{fig:vk_delta15}
    \end{subfigure}
    ~
    \begin{subfigure}[t]{0.32\textwidth}
        \includegraphics[width=\textwidth]{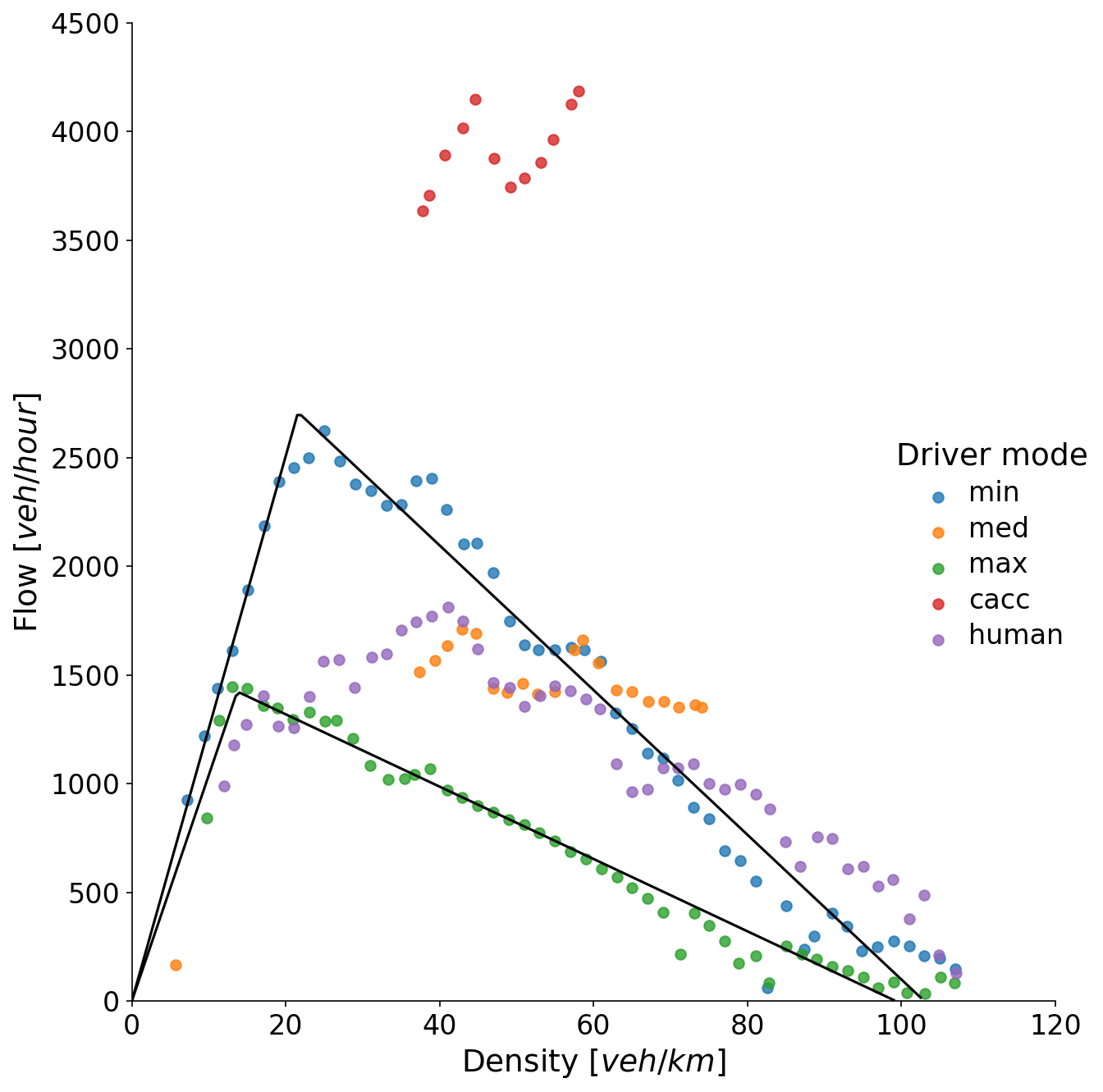}
        \caption{$q$ - $k$ plot with $\delta_{k,v}=2$.}\label{fig:vk_delta2}
    \end{subfigure}
    ~ 
    \begin{subfigure}[t]{0.32\textwidth}
        \includegraphics[width=\textwidth]{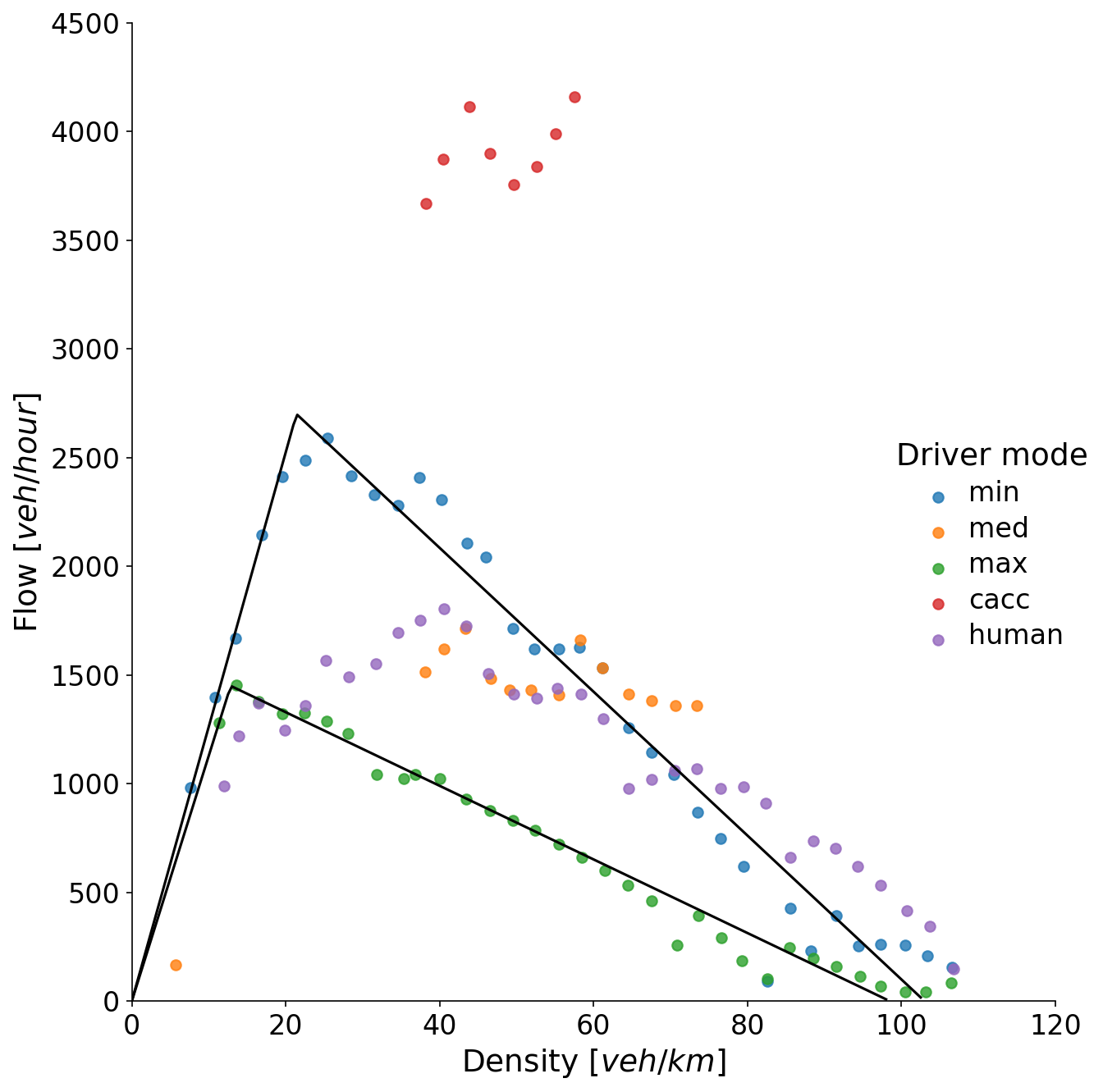}
        \caption{$v$ - $k$ plot with $\delta_{q,v}=3$.}\label{fig:vk_delta3}
    \end{subfigure}
     ~ 
    \begin{subfigure}[t]{0.32\textwidth}
        \includegraphics[width=\textwidth]{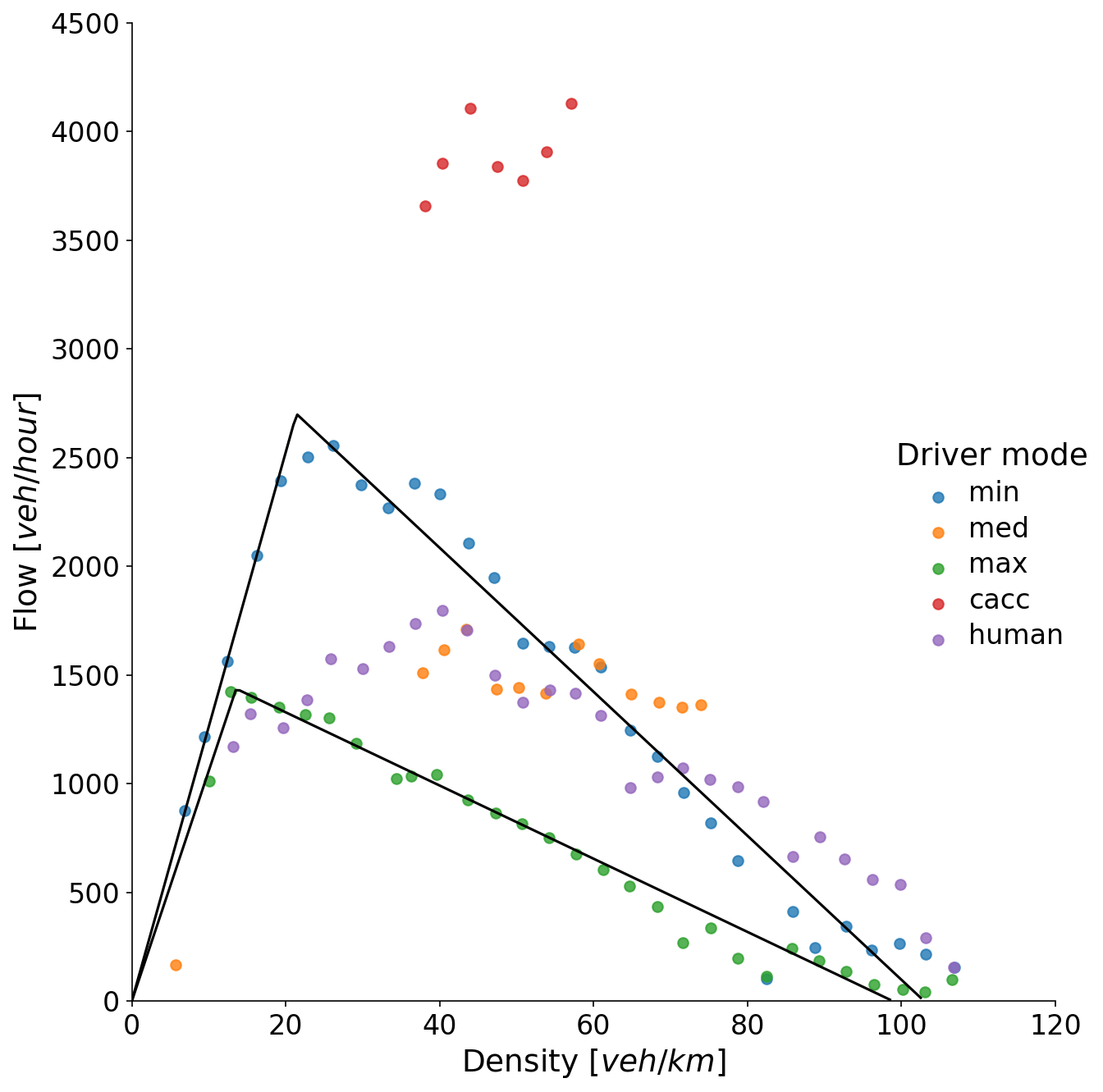}
        \caption{$q$ - $k$ plot with $\delta_{k,v}=3.5$.}\label{fig:vk_delta6}
    \end{subfigure}
    \caption{Sensitivity analysis for the consistency of the calibrated $q$ - $k$ plots of the FD for 6 additional $\delta_{k,v}$ values.} \label{fig:sensitivity}
\end{figure*}

The calibration results for different $\delta$ values are summarized in Table~\ref{tab:sensitivity}. It is interesting to observe that the calibrated TFD parameters for the minimum and maximum $\delta$s remain consistent, even for higher $\delta$ values, that normally lead to very few points for calibration. A visual illustration of the $q$ - $k$ plot for different $\delta$ values is shown in Figure~\ref{fig:sensitivity}.

\begin{table}[h]
	\caption{Calibrated TFD parameters that remain consistent for different $\delta_{k,v}$ values.}\label{tab:sensitivity}
	\begin{center} 
		\begin{tabular}{l | c c c c c c}
    		$\delta_{k,v}$ & setting & $v_f [km/h]$ & $k_{cr} [veh/km]$ & $k_{jam} [veh/km]$ & $w [km/h]$\\\hline
		 0.3~ & \textit{min} & \textcolor{blue}{126.0} & \textcolor{blue}{21.3} & \textcolor{blue}{104.4} & \textcolor{blue}{32.4}\\
		 0.3~ & \textit{max} & \textcolor{green}{110.1} & \textcolor{green}{12.9} & \textcolor{green}{101.0} & \textcolor{green}{16.2}\\ 
		 0.6~ & \textit{min} & 125.8 & 21.4 & 104.2 & 32.5\\
		 0.6~ & \textit{max} & 110.6 & 12.9 & 101.3 & 16.2\\ 
		 1.0~ & \textit{min} & 126.2 & 21.3 & 103.9 & 32.6\\
		 1.0~ & \textit{max} & 106.8 & 13.4 & 100.2 & 16.4\\ 
		 1.5~ & \textit{min} & 126.2 & 21.3 & 103.8 & 32.6\\
		 1.5~ & \textit{max} & 108.7 & 13.2 & 100.3 & 16.4\\
		 2~ & \textit{min} & 125.4 & 21.6 & 103.0  & 33.3\\
		 2~ & \textit{max} & 104.0 & 13.7 & 99.2 & 16.6\\ 
		 3~ & \textit{min} & 126.2 & 21.4 & 103.0  & 33.1\\
		 3~ & \textit{max} & 112.9 & 12.8 & 98.5 & 16.9\\ 
		 3.5~ & \textit{min} & \textcolor{blue}{126.2} & \textcolor{blue}{21.4} & \textcolor{blue}{103.0}& \textcolor{blue}{33.1}\\
		 3.5~ & \textit{max} & \textcolor{green}{106.0} & \textcolor{green}{13.6} & \textcolor{green}{98.9} & \textcolor{green}{16.8}\\
		\end{tabular}
	\end{center}
\end{table}

\clearpage

 \bibliographystyle{elsarticle-harv} 
 \bibliography{references}

\end{document}